\documentclass[a4paper,USenglish]{article} 
\usepackage{amsmath}
\usepackage{amsthm}
\usepackage{amssymb}
\usepackage{microtype}
\usepackage{complexity}
\usepackage{algorithm}			
\usepackage
	[noend]		
	{algorithmic}

\usepackage{doi}
\usepackage{xspace}
\usepackage[colorinlistoftodos]{todonotes}
\usepackage{stmaryrd} 
\usepackage{enumerate}
\usepackage{authblk}
\usepackage{subcaption}
\usepackage{tikz} 

\newtheorem{theorem}{Theorem}
\newtheorem{lemma}[theorem]{Lemma}
\newtheorem{proposition}[theorem]{Proposition}
\newtheorem{fact}[theorem]{Fact}
\newtheorem{corollary}[theorem]{Corollary}
\theoremstyle{definition}
\theoremstyle{remark}
\newtheorem{claim}{Claim}

\newcommand*{\Z}{\mathbf{Z}}

\newcommand*{\nat}{\mathbb{N}}
\newcommand*{\undirected}{\ensuremath{\mathcal{U}}}

\providecommand{\dsnp}{\textsc{Directed Steiner Forest}\xspace}

\providecommand{\krma}{\textsc{Robust $k$-Matching Augmentation}\xspace}

\newcommand*{\sets}{\ensuremath{\mathcal{S}}\xspace}
\newcommand*{\dstp}{\textsc{Directed Steiner Tree}\xspace}
\newcommand*{\nwdstp}{\textsc{Node Weighted Directed Steiner Tree}\xspace}
\newcommand*{\setcover}{\textsc{Set Cover}\xspace}

\newcommand*{\onerma}{\textsc{Robust Matching Augmentation}\xspace}

\newcommand*{\wonerpma}{\textsc{Weighted} \textsc{Robust Perfect Matching Augmentation}\xspace}
\newcommand*{\wonerma}{\textsc{Weighted} \textsc{Robust Matching Augmentation}\xspace}

\newcommand*{\ET}{\textsc{Eswaran-Tarjan}\xspace}

\newcommand*{\I}{\formatmathnames{I}\xspace}
\newcommand*{\formatmathnames}[1]{\textnormal{\small #1}}

\newcommand*{\OPT}{\formatmathnames{OPT}}
\newcommand*{\sources}{\ensuremath{V^+}\xspace}
\newcommand*{\sinks}{\ensuremath{V^-}\xspace}
\newcommand*{\tw}{\ensuremath{\mathop{tw}}\xspace}

\newcommand{\unsafe}{\text{critical}\xspace}

\newcommand{\sourcecover}{\textsc{Source Cover}\xspace}

\author[1]{Felix Hommelsheim\thanks{Research partially supported by the German Research Foundation (DFG), RTG 1855}}
\author[1]{Moritz M\"uhlenthaler$^*$}
\author[2]{Oliver Schaudt}
\affil[1]{Department of Mathematics,  TU Dortmund University}
\affil[2]{Department of Mathematics, RWTH Aachen University}

\title{How to Secure Matchings Against Edge Failures}

\usetikzlibrary{decorations.pathmorphing}
\usetikzlibrary{decorations.pathreplacing} 

\tikzset{
	edge/.style={thick, gray},
	medge/.style={decorate,very thick,decoration={snake}},
	aedge/.style={very thick,dashed,black},
	dedge/.style={thick,->},
	availedge/.style={thick,blue},
	vertex/.style={shape=circle,thick,draw,node distance=3em}
}

\begin{document} 
   
\maketitle

\begin{abstract}
    Suppose we are given a bipartite graph that admits a perfect matching and
    an adversary may delete any edge from the graph with the intention of
    destroying all perfect matchings. We consider the task of adding a minimum
    cost edge-set to the graph, such that the adversary never wins. We provide
    efficient exact and approximation algorithms. 
    In particular, for the unit-cost problem, we provide
    a $\log_2 n$-factor approximation algorithm and a
    polynomial-time algorithm for chordal-bipartite graphs. Furthermore, we
    give a fixed parameter algorithm for the problem parameterized by the treewidth of the input graph. For general
    non-negative weights we settle the approximability of the problem and show
    a close relation to the Directed Steiner Forest Problem.
    Additionally we prove a dichotomy theorem characterizing
    minor-closed graph classes which allow for a polynomial-time algorithm. Our
    methods rely on a close relationship to the classical strong connectivity
    augmentation problem and directed Steiner problems.
\end{abstract}

\newpage
\section{Introduction}

An \emph{augmentation problem} asks for a minimum-cost set of edges to be added
to a graph in order to establish a certain property. We say that a bipartite
graph is \emph{robust} if it admits a perfect matching after the removal of any
edge. Our goal is to make a bipartite graph robust at minimal cost and we study
the complexity of the corresponding augmentation problem. We refer to this
problem informally as \emph{robust matching augmentation}. As a motivation,
note that in many situations some kind of infrastructure is already available,
so we may prefer upgrading it instead of designing robust infrastructure from
scratch. Assume we have some assignment-type application, such as staff or task
scheduling, so our infrastructure is given in terms of a bipartite graph. The
application requires that we choose a perfect matching that assigns, say, tasks
to machines.  By buying additional edges, we would like to ensure that no
matter which edge fails, the resulting graph has a perfect matching, i.e., the
infrastructure remains useable. In such an application, buying edges
may correspond for example to training staff or upgrading machines. 

A complementary approach to creating robust infrastructure is captured by
design problems.  A \emph{design problem} asks for a minimum-cost subgraph with
a certain property, for instance a minimum-cost $k$-edge-connected
subgraph~\cite{cheriyan_et_al_01,gabow_et_al_kECSS_09}.
Robust matching augmentation can be stated also as a design problem, where
the given infrastructure is available at zero cost and the host graph is a
complete bipartite graph.  In fact, our problem is a special case of the
bulk-robust assignment problem, a design problem introduced
in~\cite{adjiashvili_bindewald_michaels_icalp2016}. Bulk-robustness is a
redundancy-based robustness concept proposed by Adjiashvili, Stiller and
Zenklusen~\cite{adjishvili_et_al_15}, which allows to specify a list of failure
scenarios.  The bulk-robust assignment problem is known to be \NP-hard even if
only one of two fixed edges may
fail~\cite{adjiashvili_bindewald_michaels_icalp2016}.  Here we consider the
setting that any single edge may fail.

A central theme in our algorithmic results is the occurrence  of
the classical \emph{strong connectivity augmentation} problem, which 
asks for the minimal number of arcs that are needed to make a given digraph
strongly connected.  It was shown by Eswaran and Tarjan that this problem
admits a polynomial-time algorithm, but its edge-weighted variant is \NP-hard~\cite{tarjan_augmentation_76}.
We show that also for robust matching augmentation the weighted problem is much
harder than its cardinality version. To this end, we give a $\log n$-factor
approximation algorithm for the cardinality version which is essentially tight
and prove that the weighted problem admits no $\log^{2-\varepsilon} n$-factor
approximation under standard complexity assumptions.

\paragraph*{Our Contribution}

Recall that we call a graph \emph{robust} if it admits a perfect matching
after the removal of any single edge. For a bipartite graph $(V, E)$, we denote
by $\overline{E}$ the edge-set of its bipartite complement. We provide
algorithms and hardness results for several restrictions of the following
problem.

\begin{quote}
    \onerma\\
    \textbf{instance:} Undirected bipartite graph $G=(U + W,E)$ that admits a perfect matching.\\
    \textbf{task:} Find a set $L \subseteq \overline{E}$ of minimum cardinality, such that the graph $G+L$ is robust.
\end{quote}

By a close relation of robust matching augmentation and connectivity
augmentation, we provide a deterministic $\log_2 n$-factor approximation for
\onerma, as well as a fixed parameter tractable (FPT) algorithm for the same
problem parameterized by the treewidth of the input graph. We also give a
polynomial-time algorithm for instances on chordal-bipartite graphs, which are
bipartite graphs without induced cycles of length at least six. 
Furthermore, we show that \onerma admits no polynomial-time
sublogarithmic-factor approximation algorithm unless $\P = \NP$, so our
approximation guarantee is essentially tight. 

Let us give an overview of the high-level ideas behind our algorithmic results
and make some connections to other problems. We
first show that we may restrict our attention to an arbitrary fixed perfect
matching of the input graph. That is, it suffices to prevent the adversary from
destroying a given fixed matching. From the input graph and the perfect
matching we construct an auxiliary digraph. In this digraph we select certain
sources and sinks which we connect using the Eswaran-Tarjan algorithm to obtain
a strongly connected subgraph. It turns out that strong connectivity in the
auxiliary digraph implies robustness in the original graph. We obtain an
optimal solution to our \onerma instance if the selection of sources and sinks
was optimal. 

We model the task of properly selecting sources and sinks as a variant of
the \setcover problem with some additional structure. Given an acyclic digraph, the task is to select a
minimum-cardinality subset of the sources, such that each sink is reachable
from at least one of the selected sources.  We refer to this problem
as \sourcecover and remark that its complexity may be of independent interest, since it
generalizes \textsc{Set Cover} but is a special case of \textsc{Directed
Steiner Tree}.
We give an FPT algorithm for the \sourcecover
problem parameterized by the treewidth of the input graph (ignoring
orientations).  This FPT algorithm is single exponential in the
treewidth.
As a by-product, we obtain FPT algorithms for the node-weighted and
arc-weighted versions of the \textsc{Directed Steiner Tree} problem on acyclic
digraphs, which are exponential in the treewidth and linear in the number of
nodes of the input graph.

Finally, we relax the requirement of having a perfect matching to having a
matching of cardinality at least $k$. In fact, all of our algorithmic results
for \onerma generalize to the setting where we desire to have a matching of
cardinality $k$ after deleting any single edge from a graph.

We refer by \wonerma to the generalization of \onerma, where each edge $e \in
\overline{E}$ has a non-negative cost $c_e$. The task is to find a minimum-cost set $L
\subseteq \overline{E}$, such that $G+L$ is robust.
First, we show that the approximability of \wonerma is closely linked to that of \dsnp.
In particular we show that an $f(n)$-factor approximation algorithm for \wonerma implies
an $f(n+k)$-factor  approximation algorithm for \dsnp, where $k$ is the number of terminal pairs.
By a result of Halperin and Krauthgamer~\cite{DBLP:conf/stoc/HalperinK03} it follows that there is
no $\log^{2-\varepsilon} (n)$-factor  approximation for \wonerma,
unless $\NP \subseteq \textsc{ZTIME}(n^{\polylog(n)})$.
On the positive side, we show that an
$f(k)$-factor approximation for the \dsnp problem yields an $(f(k)+1)$-factor
approximation \wonerma. Hence, the algorithms
from~\cite{CEGS08, FKN09} give an approximation guarantee of $1+
n^{\frac{1}{2}+ \varepsilon}$ for \wonerma, for every~$\varepsilon > 0$.

Second, we prove a complexity dichotomy based on
graph minors. Let $\mathcal{T}$ be a class of connected graphs closed under
connected minors. We show that \wonerma restricted to input graphs from
$\mathcal{T}$ is \NP-complete if $\mathcal{T}$ contains at least
one of two simple graph classes, which will be defined in
Section~\ref{sec:wonerpma}, and admits a polynomial-time algorithm otherwise. 
The polynomial-time algorithm for the remaining
instance classes uses a reduction to the \dsnp problem with a constant number
of terminal pairs, which in turn admits a (slice-wise) polynomial-time
algorithm due to a result by Feldman and Ruhl~\cite{feldman_ruhl_06}. The
terminal pairs of the instance are computed by the Eswaran-Tarjan
algorithm.

\paragraph*{Related work}

Adjiashvili, Bindewald and Michaels in~\cite{adjiashvili_bindewald_michaels_icalp2016} proposed an LP-based randomized
algorithm for the bulk-robust assignment problem. They
claim an $O(\log n)$-factor approximation guarantee for their algorithm.
Since the robust assignment problem generalizes \wonerma, an $O(\log n)$-factor
approximation for our problem is implied.
However, due to our inapproximability result for \wonerma, this can not be true, unless $\NP \subseteq \textsc{ZTIME}(n^{\polylog(n)})$.
The authors of~\cite{adjiashvili_bindewald_michaels_icalp2016} agree that their analysis is incorrect.

A  connectivity augmentation problem  related to strong connectivity, but
of a different flavor, is the \emph{tree augmentation problem} (TAP). The TAP
asks for a minimum-cost edge-set that increases the edge-connectivity of a
given tree from one to two. In contrast to robust matching augmentation, the
TAP admits a constant-factor approximation~\cite{frederickson_jaja_81}. The
constant has recently been lowered to $3/2 + \varepsilon$ for bounded-weight
instances~\cite{adjiashvili_tap_17,fiorini_et_al_TAP_17}. Similar to 
robust matching augmentation, the input graph is available at
zero cost. Let us briefly remark that there is more conceptual
similarity. The \emph{matching preclusion number} of a graph is the minimal
number of edges to be removed, such that the remaining graph has no
perfect matching. Robust matching augmentation can be stated as the task of
finding a minimum-cost edge-set that increases the matching preclusion number of
a bipartite graph from one to two, while the TAP aims to increase connectivity
from one to two. The matching preclusion number is considered to be a measure
of robustness of interconnect networks~\cite{brigham_et_al_05,CL:07}.
Determining the matching preclusion number of a graph is
\NP-hard~\cite{dourado_et_al_15,LMMP:12}. 

Robust perfect matchings with a given
recovery budget were studied by Dourado et al.~in~\cite{dourado_et_al_15}. Our
notion of robustness corresponds to \emph{1-robust $\infty$-recoverable} in
their terminology. They provide hardness results and structural insights mainly
for fixed recovery budgets, which bound the number of edges that can be
changed in order to repair a matching, after a certain number of edges has been
removed from the graph.

\paragraph*{Notation}
Undirected and directed graphs considered here are simple. For sets $U$, $W$,
we denote by $U + W$ their disjoint union. For an undirected bipartite graph
$G = (U+W, E)$ with bipartition $(U, W)$, we denote by $\overline{E}$ the edge-set
of its bipartite complement. Let $D = (V, A)$ be a directed graph. We
refer by $\overline A$ to the arcs not present in $D$. That is, we let $\overline{A}
\subseteq (V \times V) \setminus A$. By $\undirected (D)$ we refer to the
underlying undirected graph of $D$.
For $L \subseteq \overline{E}$, we
write $G+L$ for the graph $G'=(V(G), E(G) \cup L)$.
Simple paths in graphs are given by a sequence of vertices.  For graphs $G, H$ we
write $H \subseteq G$ if $H$ is a subgraph of $G$.  Recall that a graph  $H$ is
an \emph{induced minor} of a graph $G$ if it arises from $G$ by a sequence
of vertex deletions and edge contractions.  Similarly, the graph $H$ is a \emph{minor} of
$G$ if we additionally allow edge deletion.  Furthermore, the graph $H$ is a
\emph{connected minor} of $G$ if $H$ is connected and a minor of $G$. In
general, contractions may result in parallel edges or loops, which we simply
discard in order to keep our graphs simple.  Let $\mathcal{G}$ be a class of
graphs. We will refer to the restriction of (\textsc{Weighted}) \onerma to
instances where the graph $G$ is bipartite, admits a perfect matching, and
belongs to the class $\mathcal{G}$ as \emph{(\textsc{Weighted}) \onerma on
$\mathcal{G}$}. Given a set of items $X$ and sets $\mathcal{S} \subseteq 2^X$, the
\setcover problems asks for a minimum-cardinality subset $C \subseteq
\mathcal{S}$, such that each $x \in X$ is contained in some $s \in C$.  The
\emph{incidence graph} $G(\I)$ of a \setcover instance $\I = (X, \mathcal{S})$
is an undirected bipartite graph on the vertex set $X + \mathcal{S}$ that has
an edge $xs$ if and only if the item $x \in X$ is contained in the set $s \in
\mathcal{S}$.  

\paragraph*{Organization of the Paper}

The remainder of the paper is organized as follows. We illustrate the relation
between robust matching augmentation and strong connectivity augmentation in
Section~\ref{sec:connectivity}. Algorithms for the \sourcecover problem are
given in Section~\ref{sec:sourcecover}. Based on the results from
Sections~\ref{sec:connectivity} and~\ref{sec:sourcecover}, we present our
results on robust matching augmentation with unit costs in
Section~\ref{sec:onerma}. In Section~\ref{sec:wonerpma} we give the
complexity classification for the weighted version of the problem and
Section~\ref{sec:conclusion} concludes the paper.

\section{Robust Matchings and Strong Connectivity Augmentation}
\label{sec:connectivity}

In this section we give some preliminary observations on the close relationship
between robust matching augmentation with unit costs and strong connectivity
augmentation. For this purpose, we fix an arbitrary perfect matching and
construct an auxiliary digraph that is somewhat similar to the
\emph{alternating tree} used in Edmond's blossom algorithm. We show that the
original graph is robust if the auxiliary graph is strongly connected (but not
vice versa). Furthermore, we show that there is an optimal edge-set making the
given graph robust, that corresponds to a set of arcs connecting sources and
sinks in the auxiliary digraph. Finally, if no source or sink of the auxiliary
digraph corresponds to a non-trivial robust part of the original graph, then we
may use the algorithm for strong connectivity augmentation by Eswaran and
Tarjan~\cite{tarjan_augmentation_76} to make the original graph robust. As a
consequence, we have that \onerma on trees can be solved efficiently by using
the Eswaran-Tarjan algorithm. In Section~\ref{sec:onerma}, we will generalize
this result.

Let $G = (U + W, E)$ be a bipartite graph that admits a perfect
matching and let $M$ be an arbitrary but fixed perfect matching $M$ of $G$. We
call an edge $e \in M$ \emph{\unsafe} if $G - e$ admits no perfect matching.
Observe that an edge $e \in M$ is \unsafe if and only if it is not contained in
an $M$-alternating cycle. Furthermore, no edge in $E \setminus M$ is \unsafe.
Since $M$ is perfect, each edge $e \in M$ is incident to a unique vertex $u_e$
of $U$. 
We consider the following auxiliary digraph $D(G, M) = (U, A)$, whose arc-set
$A$ is given by
\begin{align*}
  A :=& \{ uu' \mid u, u' \in U : \text{there is a vertex } w \in W \text{ such that } uw \in M \text{ and } wu' \in E \setminus M \}
.
\end{align*}
We first note that the choice of the bipartion of $G$ is irrelevant. 
\begin{fact}
  \label{fact:iso}
  Let $G' = (U' + W', E)$, where $(U', W')$ is a bipartition of $G$. Then $D(G, M)$ is isomorphic to $D(G', M)$.
\end{fact}

Note that we may perform the reverse construction as well. That is, from any
digraph $D'$ we may obtain a corresponding undirected graph $G$ and a perfect
matching $M$ of $G$ such that $D(G, M) = D'$. In fact, augmenting edges to $G$
is equivalent to augmenting arcs to $D(G, M)$.

\begin{fact}
    \label{fact:arc-edge-correspondence}
    Let $\overline{A}$ be the set of arcs that are not present in $D(G,
    M)$. Then there is a 1-to-1 correspondence between $\overline{E}$ and
    $\overline{A}$.
\end{fact}

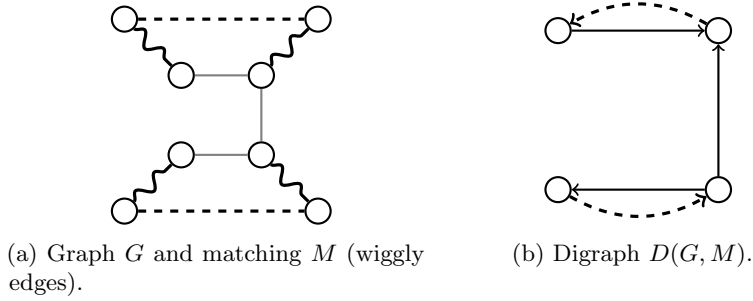
\begin{figure}[t]
  \centering
\subcaptionbox{Graph $G$ and matching $M$ (wiggly edges).\label{fig:augmentation:graph}}[0.45\textwidth]{
	\begin{tikzpicture}
	  \node[vertex] (v1) {};
	  \node[vertex,below right of=v1] (v2) {};
	  \node[vertex,right of=v2] (v3) {};
	  \node[vertex,above right of=v3] (v4) {};

	  \node[vertex,below of=v3] (w3) {};
	  \node[vertex,left of=w3] (w2) {};
	  \node[vertex,below left of=w2] (w1) {};
	  \node[vertex,below right of=w3] (w4) {};

	  \draw[medge] (v1) -- (v2);
	  \draw[medge] (v3) -- (v4);
	  \draw[medge] (w1) -- (w2);
	  \draw[medge] (w3) -- (w4);
	  \draw[edge] (v2) -- (v3) -- (w3) -- (w2);

	  \draw[aedge] (v1) -- (v4);
	  \draw[aedge] (w1) -- (w4);
	\end{tikzpicture}
  }
  \hspace{2em}
  \subcaptionbox{Digraph $D(G, M)$.\label{fig:augmentation:digraph}}[0.30\textwidth]{
	\begin{tikzpicture}[vertex/.append style={node distance=6em}]
	  \node[vertex] (v1) {};
	  \node[vertex,right of=v1] (v2) {};
	  \node[vertex,below of=v2] (w2) {};
	  \node[vertex,left of=w2] (w1) {};

	  \draw[dedge] (v1) -- (v2);
	  \draw[dedge] (w2) -- (v2);
	  \draw[dedge] (w2) -- (w1);

	  \draw[aedge,->] (w1) to [bend right=30] (w2);
	  \draw[aedge,->] (v2) to [bend right=30] (v1) {};
	\end{tikzpicture}
  }
  \caption{Illustration of the correspondence between the dotted edges in
   and 
   .} 
  \label{fig:augmentation:example}
\end{figure}

An example of the correspondence mentioned in
Fact~\ref{fact:arc-edge-correspondence} is shown in
Figure~\ref{fig:augmentation:example}.  In order to keep our notation tidy, we
will make implicit use of Fact~\ref{fact:arc-edge-correspondence} and refer to
$\overline{A}$ and $\overline{E}$ interchangeably. Observe that for edges $e,
f \in M$ there is an $M$-alternating path containing $e$ and $f$ in $G$ if and
only if $u_e$ is connected to $u_f$ in $D(G, M)$. This implies the following
characterization of robustness.

\begin{fact}
  \label{fact:kstrong}
  $G$ is robust if and only if each strongly connected component of
  $D(G, M)$ is non-trivial, that is, it contains at least two vertices.
\end{fact}

Let $D'$ be a digraph. A vertex of $D'$ is called a \emph{source}
(\emph{sink}) if it has no incoming (outgoing) arc. We refer to the set of
sources (sinks) of $D'$ by $\sources(D')$ ($\sinks(D')$). Furthermore, we
denote by $C(D')$ the \emph{condensation} of  $D'$, that is, the directed
acyclic graph of strongly connected components of $D'$. We call a source or
sink of $C(D')$ \emph{strong} if the corresponding strongly connected component
of $D'$ is non-trivial.
From Fact~\ref{fact:kstrong} it follows that a subgraph of $G$ that
corresponds to a strong source or a strong sink is robust against the failure
of a single edge. Furthermore, observe that the choice of the perfect matching
$M$ of $G$ is irrelevant in the following sense.

\begin{fact}
  \label{fact:iso2}
  Let $M$ and $M'$ be perfect matchings of $G$. Then $C(D(G, M))$ is
  isomorphic to $C(D(G, M'))$. 
\end{fact}

Fact~\ref{fact:iso2} is of key importance for our algorithmic results, for
which we generally assume that some fixed perfect matching is given. Next, we
observe that for unit costs we may restrict our attention to connecting
sources and sinks of $C(D)$ in order to make $G$ robust. It is easy to check
that this does not hold for general non-negative costs.

\begin{fact}
    \label{fact:sourcesink}
    Let $L \subseteq \overline{E}$ such that $G + L$ is robust. Then there is
    some $L' \subseteq \overline{E}$ of cardinality at most $|L|$, such that $G
    + L'$ is robust and $L'$ connects only sinks to sources of $C(D(G, M))$. 
\end{fact}
We remark that the construction of $L'$ given in the proof of
Fact~\ref{fact:sourcesink} can be performed in polynomial time. 

We denote by $\gamma(D')$ the minimal number of arcs to be added to a
digraph $D'$ in order to make it strongly connected. Eswaran an Tarjan have
proved the following min-max relation~\cite{tarjan_augmentation_76}.
\begin{fact}
  \label{fact:minmax}
  Let $D'$ be a digraph. Then $\gamma(D') = \max \{ |\sources(D')|, |\sinks(D')| \}$.
\end{fact}

From the proof of Fact~\ref{fact:minmax} it is easy to obtain a polynomial-time
algorithm that, given a digraph $D'$, computes an arc-set $L$ of cardinality
$\gamma(D')$ such that $D' + L$ is strongly
connected~\cite{frank_jordan_augmentation}. We will refer to this algorithm by
\ET.  The following proposition illustrates the usefulness of the algorithm \ET
for \onerma, and at the same time its limitations. 

\begin{fact}
  \label{fact:eswaran-tarjan-and-optimal-solutions}
  Suppose that $C(D(G, M))$ contains no strong sources or sinks. Then \ET computes a
  set $L \subseteq \overline{E}$ of minimum cardinality such that $G + L$ is robust. 
\end{fact}

Fact~\ref{fact:eswaran-tarjan-and-optimal-solutions} implies that \ET solves
\onerma on trees.  If strong sources or sinks are present in $D(G, M)$, then we
may or may not need to consider them in order to make $G$ robust. This is
precisely what makes the problem \onerma hard. We will formalize the task of
selecting strong sources and sinks in terms of the \sourcecover problem, which
is discussed in the next section.

\section{The \sourcecover Problem}
\label{sec:sourcecover}

To present our algorithmic results in Section~\ref{sec:onerma} in a concise
fashion it will be convenient to introduce the \sourcecover
problem. Given an acyclic digraph, the \sourcecover problem asks for a
minimum-cardinality subset of its sources, such that each sink is reachable
from at least one selected source.  It is easy to see that \sourcecover is a
special case of the \dstp problem and that it generalizes \setcover. We give a
simple polynomial-time algorithm for \sourcecover if the input graph is
chordal-bipartite (ignoring orientations). Furthermore, we show that
\sourcecover parameterized by treewidth (again ignoring
orientations) is FPT. As a by-product, we obtain a simple FPT algorithm for the
arc-weighted and node-weighted versions of the \textsc{Directed Steiner Tree}
problem on acyclic digraphs, whose running time is linear in the size of the
input graph and exponential in the treewidth of the underlying undirected
graph. To the best of our knowledge, the parameterized complexity of the
general \textsc{Directed Steiner Tree} problem with respect to treewidth is
open.  For the corresponding \emph{undirected} \textsc{Steiner Tree} problem,
an FPT algorithm was given by Bodlaender et al.~in~\cite{BCKN:15}.

The \sourcecover problem is formally defined as follows.

\begin{quote}
  \textsc{\sourcecover}\\
  \textbf{instance:} Weakly connected acyclic digraph $D = (V, A)$ with at least one arc.\\
  \textbf{task:} Find a minimum-cardinality subset $S$ of the sources $\sources(D)$ of $D$, such that for each sink $t \in \sinks(D)$ there is an $S$-$t$-path in $D$.
\end{quote}

The assumptions that $D$ is connected and contains at least one arc are present
only for technical reasons.
By ``flattening'' the input digraph, we can turn an instance $\I = (D)$ of
\sourcecover into a \setcover instance as follows.
Let $F(D) = (\sources(D) \cup \sinks(D), A')$ be an acyclic digraph, where $A'$ is
given by 
\[
  A' := \{ st \mid s \in \sources(D),\, t \in \sinks(D),\, \text{$t$ is reachable from $s$ in $D$} \} .
\]
Then $\undirected(F(D))$ is the incidence graph of a \setcover instance
$\mathcal{A}$ on $\sinks(F(D))$, such that the feasible solutions of \I and
$\mathcal{A}$ are in 1-to-1 correspondence. 

\begin{figure}[t]
  \centering
  \subcaptionbox{A digraph $D$ such that $\undirected(D)$ is balanced, but $\undirected(F(D))$ is not.\label{fig:balanced}}[0.45\textwidth]{
    \begin{tikzpicture}
      \node[vertex] (u1) {};
      \node[vertex,above of=u1] (v1) {};
      \node[vertex,above of=v1] (w1) {};

      \node[vertex, right of=u1] (c1) {};
      \node[vertex, right of=w1] (c2) {};

      \node[vertex,right of=c1] (x1) {};
      \node[vertex,above of=x1] (y1) {};
      \node[vertex,above of=y1] (z1) {};

      \draw[edge,->] (u1) -- (z1);
      \draw[edge,->] (u1) -- (c1);
      \draw[edge,->] (v1) -- (y1);
      \draw[edge,->] (v1) -- (x1);
      \draw[edge,->] (w1) -- (c2);
      \draw[edge,->] (w1) -- (y1);
      \draw[edge,->] (c1) -- (x1);
      \draw[edge,->] (c2) -- (z1);
    \end{tikzpicture}
  }
  \hspace{2em}
  \subcaptionbox{Digraphs $D$ such that $\undirected(D)$ has
    treewidth one, but the treewidth of $\undirected(F(D))$ is unbounded.\label{fig:treewidth}}[0.45\textwidth]{
    \begin{tikzpicture}
      \node[vertex] (c1) {};

      \node[vertex,right of=c1]  (v1) {};
      \node[vertex,draw=none,above of=v1] (rdots) {$\vdots$};
      \node[vertex,above of=rdots] (u1) {};
      \node[vertex,below of=v1] (w1) {};

      \node[vertex,left of=c1]   (x1) {};
      \node[vertex,draw=none,above of=x1] (ldots) {$\vdots$};
      \node[vertex,above of=ldots] (y1) {};
      \node[vertex,below of=x1] (z1) {};

      \draw[edge,->] (x1) -- (c1);
      \draw[edge,->] (y1) -- (c1);
      \draw[edge,->] (z1) -- (c1);
      \draw[edge,->] (c1) -- (v1);
      \draw[edge,->] (c1) -- (w1);
      \draw[edge,->] (c1) -- (u1);
    \end{tikzpicture}
  }
  \caption{Examples showing that flattening does not preserve balancedness or bounded treewidth.}
  \label{fig:counter}
\end{figure}
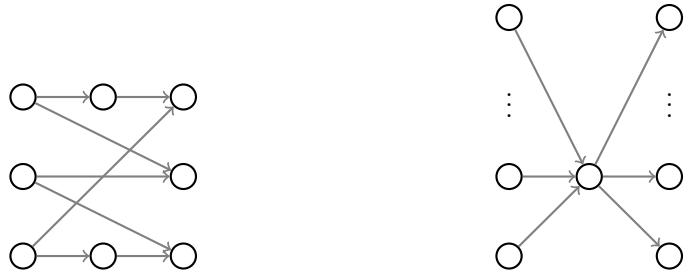

As illustrated in Figure~\ref{fig:counter}, useful properties of the input
digraph may not be present in the corresponding flattened digraph. In
particular, if $\undirected(D)$ has treewidth at most $r$, then the treewidth
of $\undirected(F(D))$ cannot be bounded by a constant in general. Furthermore,
the graph $\undirected(F(D))$ is not necessarily balanced\footnote{A graph is called \emph{balanced} if the length of each induced cycle is divisible by four.} (or planar) if
$\undirected(D)$ is. Therefore, we cannot take advantage of polynomial-time
algorithms for \setcover on balanced incidence graphs or incidence graphs of
bounded treewidth. 
Motivated by the example in Figure~\ref{fig:treewidth} we leave as an open
question, whether \sourcecover on balanced graphs admit polynomial-time
algorithms. By Theorem~\ref{lemma:main}, the existence of such an algorithm
implies a polynomial-time algorithm for \onerma on balanced graphs. 

\subsection{\sourcecover on Chordal Bipartite Graphs}

We show that in contrast to the treewidth and balancedness, chordal-bipartiteness is indeed preserved by the flattening operation introduced above.
From this we obtain the following result.

\begin{theorem}
  \sourcecover on chordal-bipartite graphs admits a polynomial-time algorithm.
  \label{thm:sc:chordal-bipartite}
\end{theorem}

To prove the theorem, we show that if $\undirected(D)$ is chordal-bipartite,
so is $\undirected(F(D))$. The graph $\undirected(F(D))$ is the incidence
graph of a \setcover instance, whose optimal solutions correspond canonically
to the optimal solutions of the \sourcecover instance $(D)$. It is known that
\setcover on chordal-bipartite incidence graphs  (and more generally, balanced
graphs) admits a polynomial-time algorithm: 
It is possible to use
LP-methods and the fact that covering polyhedra of balanced matrices are
integral, see~\cite[pp.~562-573]{nemhauser_wolsey}. Alternatively we can use
a combinatorial algorithm by Hoffman et al.~\cite{HKS:85}.

\subsection{\sourcecover on Graphs of Bounded Treewidth}
\label{subsec:tw}

We provide a fixed-parameter algorithm for \nwdstp on acyclic digraphs that is
single-exponential in the treewidth of the underlying undirected graph and
linear in the instance size. Since \sourcecover is a restriction of \nwdstp on acyclic graphs,
we have a polynomial-time algorithm for \sourcecover parameterized by the
treewidth of the underlying undirected graph. Let us first recall some
definitions related to Steiner problems and tree decompositions.

\begin{quote}
  \textsc{\nwdstp}\\
  \textbf{instance:} Acyclic digraph $D = (V, A)$, costs $c \in \mathbb{R}_{\geq 0}^V$, terminals $T \subseteq V$, root $r \in V$.\\
  \textbf{task:} Find a minimum-cost subset $F \subseteq V$, such that $r$ is connected to each terminal in $(F, E(F))$.
\end{quote}

\textsc{Arc Weighted Directed Steiner Tree} is the corresponding problem, where
the costs are on the arcs of the graph.
A \emph{tree decomposition} of a graph $G = (V, E)$ is a tree $T$ as follows.
Each node $x \in V(T)$ of $T$ has a \emph{bag} $B_x \subseteq V$ of vertices of
$G$ such that the following properties hold.
\begin{itemize}
    \item $\bigcup_{x \in V(T)} B_x = V$.
    \item If $B_x$ and $B_y$ both contain a vertex $v \in V$, then the bags of all nodes of $T$ in the path between $x$ and $y$ contain $v$ as well. Equivalently, the tree nodes containing vertex $v$ form a connected subtree of $T$.
    \item For each edge $vw$ in $G$ there is some bag that contains both $v$ and $w$. That is, for vertices adjacent in $G$, the corresponding subtrees have a node in common.
\end{itemize}
The \emph{width} of a tree decomposition is the size of its largest bag minus
one. The \emph{treewidth} $\tw(G)$ of $G$ is the minimum width among all
possible tree decompositions of $G$.

To solve the \nwdstp on acyclic digraphs, we use a simple dynamic-programming
algorithm over the tree decomposition of the underlying undirected graph of the
input digraph $D$ with $n$ vertices.

\begin{theorem}
  \nwdstp on acyclic digraphs can be solved in time $O(5^{w} \cdot w \cdot n)$ if
  a tree decomposition of $\undirected(D)$ of width $w$ is provided.
  \label{thm:dstp:tw}
\end{theorem}

Note that an optimal tree-decomposition of a graph $G$ can be computed in time
$O(2^{O(\tw(G)^3)}\cdot n)$ by Bodlaender's famous
theorem~\cite{DBLP:journals/siamcomp/Bodlaender96}.  Our algorithm intuitively
works in the following way and is similar to the dynamic programming algorithm
for \textsc{Dominating Set} (see,
e.g.,~\cite[Section~7.3.2]{param_complexity_book_15}).  We interpret a solution
to \nwdstp as follows: each vertex of $D$ may be \emph{active} or not.  Each
active vertex needs a predecessor that is also active, unless it is the root.
The cost to activate a vertex is given by the cost function of the \nwdstp
instance.  Starting with all terminals active, it is easy to see that \nwdstp
on acyclic graphs is equivalent to the problem of finding a minimum cost active
vertex set satisfying the above conditions.  
We compute an optimal solution in a bottom-up
fashion using a so-called \emph{nice} tree decomposition of the input graph.

By a simple reduction, we also obtain an \FPT-time algorithm for \textsc{Arc
Weighted Directed Steiner Tree} on acyclic digraphs. We just subdivide each arc
and assign the cost of the arc to the corresponding new vertex. Each old
vertex receives cost zero. This transformation does not increase the treewidth.

Furthermore, we can reduce \sourcecover to \nwdstp by adding a new vertex $r$ and connecting
$r$ to each source by an arc. The sources have cost one, while all other vertices have cost zero. 
The root vertex is $r$ and the set of terminals is the set of sinks. 
By adding only one new vertex, the treewidth is increased by at most one.
As a consequence of this reduction and Theorem~\ref{thm:dstp:tw}, we obtain the
following result.

\begin{corollary}
  \sourcecover can be solved in time $O(5^{w}\cdot w \cdot n)$ if a tree-decomposition of $\undirected(D)$ of width $w$ is provided.
  \label{thm:sc:tw}
\end{corollary}

\section{Robust Matching Augmentation}
\label{sec:onerma}

In this section we present our main results on the problem \onerma. Let us
first redefine the problem in a slightly different way.

\begin{quote}
    \onerma\\
    \textbf{instance:} Bipartite graph $G = (U + W, E)$ and perfect matching $M$ of $G$.\\
    \textbf{task:} Find a minimum-cardinality set $L \subseteq \overline{E}$ such that $G + L$ is robust.
\end{quote}

Fixing the perfect matching $M$ in the instance is just for notational 
convenience, since we can compute a perfect matching in polynomial time and our
results do not depend on the exact choice of $M$, according to the discussion in
Section~\ref{sec:connectivity}. The next theorem is our main technical result of
this section. By combining the theorem with the results in
Section~\ref{sec:sourcecover} we obtain our algorithmic results.

\begin{theorem}
  There is a polynomial-time algorithm that, given an instance $\I = (G, M)$ of
  \onerma, computes two instances $\mathcal{A}_1 = (S_1)$ and $\mathcal{A}_2 =
  (S_2)$ of \sourcecover such that the following holds.
  \begin{enumerate}
    \item $\undirected(S_1)$ and $\undirected(S_2)$ are induced minors of $\undirected(D(G, M))$.
    \item $\OPT(\I) = \max \{ \OPT(\mathcal{A}_1), \OPT(\mathcal{A}_2) \}$ \label{itm:main:opt}
    \item From a solution $C_1$ of $\mathcal{A}_1$ and a solution $C_2$ of
      $\mathcal{A}_2$ we can construct in polynomial time a solution $L$ of \I
      of cardinality $\max \{|C_1|, |C_2|\}$. \label{itm:main:solution}
  \end{enumerate}
  \label{lemma:main}
\end{theorem}
\begin{proof}
  Let $\I = (G, M)$ be an instance of \onerma, where $G = (U + W, E)$.  Our
  goal is to obtain from solutions of the \sourcecover instances  a suitable
  selection of sources and sinks of $C(D(G, M))$, such that we can make $M$
  robust by connecting the selected sources and sinks, using the
  algorithm \ET.  Let us denote by $u_e$ the vertex in $U$ that is incident to
  an edge $e \in M$. Furthermore, let $D := D(G, M)$. We construct the
  \sourcecover instance $\mathcal{A}_1$ as follows. For each \unsafe edge
  $e \in M$, we remove from $D$ each vertex $v \in U - u_e$, such that $v$ is
  reachable from $u_e$ in $D$. Let $D'$ be the resulting graph and let the
  \sourcecover instance $\mathcal{A}_1$ be given by  $\mathcal{A}_1 :=
  (C(D'))$. The construction of
  $\mathcal{A}_2$ is as for $\mathcal{A}_1$, but with the arcs of $D$ reversed.
  This turns the sources of $D$ into sinks. 
  Clearly, the acyclic digraphs of $\mathcal{A}_1$ and $\mathcal{A}_2$ are
  induced minors of $\undirected(D)$, since they were constructed by deleting
  vertices of $\undirected(D)$ and contracting strong components.  By
  Fact~\ref{fact:kstrong}, the set of \unsafe edges can be obtained
  efficiently by Tarjan's classical algorithm for computing strongly connected
  components. In order to generate $\mathcal{A}_1$ and $\mathcal{A}_2$, observe
  that $D'$ and $C(D')$ can both be obtained by applying a breadth-first search
  starting at each vertex of $D$ or $D'$, respectively.   So it remains to
  prove Statement~\ref{itm:main:opt} and~\ref{itm:main:solution}.

  Let $C_1$ ($C_2$) be a solution to $\mathcal{A}_1$ ($\mathcal{A}_2$). We show
  how to construct in polynomial time a solution $L$ of $\I$ of cardinality
  $\max \{|C_1|, |C_2|\}$. Let $X \subseteq V(\widehat D)$ be the set of vertices incident to
  \unsafe edges. Moreover, let $\widehat D \subseteq C(D)$ be the graph induced by the
  vertices of $C(D)$ that are on $C_1X$-paths or on $XC_2$-paths in $C(D)$.
  Note that $\widehat D$ can be computed by a depth-first search
  applied on each source and sink.
  By running \ET on $\widehat D$ we obtain an arc-set
  $L^*$ such that $\widehat D + L^*$ is strongly connected. Hence, each $u \in X$ is
  on some directed cycle in $\widehat D + L^*$. From $L^*$ we can obtain in a
  straight-forward way an arc-set $L$ of the same cardinality, such that each
  $u \in X$ is on some directed cycle of $D + L$. For each $ss' \in L^*$, we
  add to $L$ an arc $uu'$, where $u$ ($u'$) is some vertex in the strong
  component $s$ ($s'$) of $D$. By the construction of $L$, each $u \in X$ is on
  some directed cycle of $D$. By Fact~\ref{fact:arc-edge-correspondence}
  and~\ref{fact:minmax} we have constructed a solution $L$ of $\I$ of
  cardinality $|L| = |L^*| = \max \{|C_1|, |C_2| \}$. 
  This completes the proof
  of Statement~\ref{itm:main:solution}.

  It remains to prove that $\OPT(\I) \geq \max \{ \OPT(\mathcal{A}_1),
  \OPT(\mathcal{A}_2) \}$. Suppose for a contradiction that $\OPT(\I) <
  \max\{\OPT(\mathcal{A}_1), \OPT(\mathcal{A}_2)\}$. Without loss of
  generality, let $\OPT(\mathcal{A}_1)$ attain the maximum.
  Due to Fact~\ref{fact:sourcesink}, we may assume that an optimal solution $L$
  of $\I$ connects sources and sinks of $C(D)$. Let $R \subseteq V(C(D))$
  be the corresponding sources of $C(D)$. Then for each \unsafe edge $e \in
  M$, the vertex $u_e$ must be reachable from some source $s \in R$.  But
  then $R$ is a solution of $\mathcal{A}_1$ of cardinality $|R| = \OPT(\I)
  < \OPT(\mathcal{A}_1)$, a contradiction.
\end{proof}

As a first consequence of Theorem~\ref{lemma:main} we obtain a simple
$\log_2 n$-factor approximation algorithm for \onerma.
We ``flatten'' the graph of the \textsc{Source Cover} instances as described in
Section~\ref{sec:sourcecover} to obtain \setcover instances and then use
the classic Greedy-Algorithm to achieve a $\log_2 n$-factor approximation.

\begin{corollary}
  \onerma admits a polynomial-time $\log_2 n$-factor
  approximation algorithm, where $n$ is the number of vertices of the input
  graph. 
  \label{cor:onerma:approx}
\end{corollary}

In a similar fashion we obtain a polynomial-time algorithm on chordal-bipartite
graphs by combining Theorems~\ref{lemma:main}
and~\ref{thm:sc:chordal-bipartite} and the observation that $\undirected(D(G,
M))$ is chordal-bipartite if $G$ is. Furthermore, we give an FPT algorithm
parameterized by the treewidth by combining Theorems~\ref{lemma:main} and
Corollary~\ref{thm:sc:tw} and the observation that treewidth is monotone under
taking minors.

\begin{corollary}
    \onerma admits a polynomial-time algorithm on chordal-bipartite graphs and
    an FPT algorithm parameterized by the treewidth of the input graph.
    \label{cor:onerma:tw-cb}
\end{corollary}

We now show that our algorithms are also applicable in the
following more general setting. Suppose we would like to have a matching of a
given cardinality in the graph, no matter which edge is deleted by the
adversary.  
\begin{quote}
  \krma\\
  \textbf{instance:} Bipartite graph $G = (U + W, E)$ that admits a matching of size $k$.\\
  \textbf{task:} Find a minimum-cardinality set $L \subseteq \overline{E}$ such that for $e \in E$, the graph $G + L - e$ admits a matching of size $k$.
\end{quote}

Note that if $k$ is not the size of a maximum matching, then $L = \emptyset$ is
feasible
due to
the existence of a larger matching. We give a polynomial-time reduction from
\krma to \onerma that increases the treewidth by at most two.
On the other hand, chordal-bipartiteness of the input graph is not preserved 
However, the corresponding digraph contains no induced cycle of
length at least six, so Theorem~\ref{thm:sc:chordal-bipartite} is still
applicable. By Proposition~\ref{thm:onerma:reduction} and the previous
corollaries, we obtain for \krma a $\log_2 n$-factor approximation algorithm, a
polynomial-time algorithm on chordal-bipartite graphs, and an FPT algorithm
parameterized by the treewidth.

\begin{proposition}
    There is a polynomial-time reduction $f$ from  \krma to \onerma, such that
    the following holds.  Let $\I := (G)$ be an instance of \krma and let $f(\I) =
    (G')$. Then
    \begin{enumerate}
    \item $\OPT(f(\I)) = \OPT(\I)$ and from a solution $L'$ of $f(\I)$ 
  we can construct in polynomial time a solution $L$ of $\I$ such that $|L| \leq |L'|$.\label{itm:onerma:max:perf}
    \item $\tw(G') \leq \tw(G) + 2$\label{itm:onerma:tw}
    \item If $G$ is chordal-bipartite then $\undirected(D(G', M'))$ has no induced cycle of length at least six.\label{itm:onerma:no induced cycle on length six}
  \end{enumerate}
  \label{thm:onerma:reduction}
\end{proposition}

\section{Weighted Robust Matching Augmentation}
\label{sec:wonerpma}

As shown above, \onerma is tightly linked to \setcover in terms of approximation.
Our first result in this section shows that \wonerma is substantially more complicated, as its approximability is closely linked to \dsnp. This problem is formally defined as follows:
\begin{quote}
	\dsnp\\
	\textbf{instance:} Directed graph $G = (V, A)$, $k$ terminal pairs $(s_i, t_i)_{1 \leq i \leq k}$, costs $w \in \Z_{\geq 0}^A$.\\
	\textbf{task:} Find a minimum-cost subgraph $G' \subseteq G$ such that for each $1 \leq i \leq k$, the vertex $s_i$ is connected to $t_i$ in $G'$.
\end{quote}

\begin{proposition}
\label{prop:hardness:onerma-dsf}
Let $n'$ be the number of vertices of the \wonerma instance and $n$ and $k$ 
be the number of vertices and terminals of the \dsnp instance, respectively.

An $f(n')$-factor approximation algorithm for \wonerma implies an 
$f(n+k)$-factor approximation algorithm for \dsnp.
An $f(n)$- or an $f(k)$-factor approximation algorithm for \dsnp imply an $(f(n)+1)$- or
$(f(k)+1)$-factor approximation algorithm for \wonerma, respectively.
\end{proposition}

On the one hand this result implies an $(n^{1/2+\varepsilon} + 1)$-factor approximation algorithm 
for \wonerma
for every $\varepsilon > 0$, due to~\cite{CEGS08, FKN09}, who achieve
a guarantee of $k^{1/2 + \varepsilon}$, for every~$\varepsilon > 0$.
On the other hand, an algorithm achieving a guarantee of $n^{1/3}$ or 
better for \wonerma implies a better
approximation algorithm for \dsnp, as the number 
$k$ of distinct terminal pairs is at most $O(n^2)$ and 
the current best approximation factor in terms of $n$ 
is $n^{2/3 + \varepsilon}$ due to Berman et al.~\cite{BBMRY13}.
Additionally, by a result of Halperin and Krauthgamer~\cite{DBLP:conf/stoc/HalperinK03}, the above proposition implies the following lower bound.

\begin{corollary}
  \label{cor:wonerma:dsf:hardnessOfApproximation}
  For every $\varepsilon > 0$ \wonerma does not admit a $\log^{2- \varepsilon}(n)$-factor approximation algorithm unless $\NP \subseteq \textsc{ZTIME}(n^{\polylog(n)})$. 
\end{corollary}

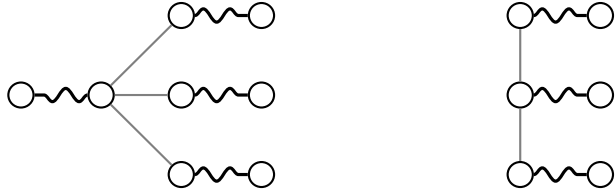
\begin{figure}[t]
	\centering
		\begin{tikzpicture}
		\node[vertex] (u1) {};
		\node[vertex,right of=u1] (u2) {};
		\node[vertex,above of=u1] (v1) {};
		\node[vertex,right of=v1] (v2) {};
		\node[vertex,above of=v1] (w1) {};
		\node[vertex,right of=w1] (w2) {};
		\node[vertex,left of =v1] (c2) {};
		\node[vertex,left of =c2] (c1) {};
		
		\draw[medge] (c2) -- (c1);
		\draw[medge] (u1) -- (u2);
		\draw[medge] (v1) -- (v2);
		\draw[medge] (w1) -- (w2);
		\draw[edge]  (c2) -- (u1);
		\draw[edge]  (c2) -- (v1);
		\draw[edge]  (c2) -- (w1);
		\end{tikzpicture}
	\hspace{8em}
		\begin{tikzpicture}
		\node[vertex] (u1) {};
		\node[vertex,right of=u1] (u2) {};
		\node[vertex,above of=u1] (v1) {};
		\node[vertex,right of=v1] (v2) {};
		\node[vertex,above of=v1] (w1) {};
		\node[vertex,right of=w1] (w2) {};
		
		\draw[medge] (u1) -- (u2);
		\draw[medge] (v1) -- (v2);
		\draw[medge] (w1) -- (w2);
		\draw[edge]  (u1) -- (v1) -- (w1);
		\end{tikzpicture}
	\caption{The graphs $K_{1,3}^*$ and $P_3^*$, each with its unique perfect matching.}
	\label{fig:K1r-Pr}
\end{figure}

Given this negative result we proceed to the analysis of structural restrictions that make \wonerma more accessible.
The main result of this section is a classification of the complexity of the
problem \wonerma on minor-closed graph classes. In particular we show that the problem is
\NP-hard on a minor-closed class $\mathcal{G}$ of graphs if and only if $\mathcal{G}$
contains at least one of the two graph classes $\mathcal{K}^*$ and
$\mathcal{P}^*$, which we will define next.  Let $K_{1,r}$ be the star graph
with $r$ leaves and let $P_r$ be the path on $r$ vertices. For any graph $H$
let $H^*$ be the graph obtained by attaching a leaf to each vertex of $H$. Then
$\mathcal{K}^* := \{ K_{1,r}^* \mid r \in \nat\}$ and $\mathcal{P}^* := \{P_r^*
\mid r \in \nat\}$.  Note that each graph in $\mathcal{K}^*$ and
$\mathcal{P}^*$ has a unique perfect matching.  See Figure~\ref{fig:K1r-Pr} for
an illustration of the graphs $K_{1,3}^*$ and $P_{3}^*$.

\begin{lemma}
	\label{lemma:hardness:stars-and-paths}
	\wonerma is \NP-hard on each of the classes $\mathcal{K}^*$ and $\mathcal{P}^*$.
\end{lemma}

We complement Lemma~\ref{lemma:hardness:stars-and-paths} by showing that \wonerma
on a class $\mathcal{G}$ of graphs admits a polynomial-time algorithm if
$\mathcal{G}$ contains neither $\mathcal{K}^*$ nor $\mathcal{P}^*$.

\begin{theorem}
	\label{thm:weighted-main}
	Let $\mathcal G$ be a class of connected graphs that is closed under connected minors.
	Then \wonerma on $\mathcal{G}$ admits a polynomial-time algorithm if and only
	if there is some $r \in \nat$ such that $\mathcal G$ contains neither
	the graph $K_{1,r}^*$ nor $P_r^*$. The only if part holds under the assumption that $\P
	\neq \NP$.
\end{theorem}

In order to prove Lemma \ref{lemma:hardness:stars-and-paths}, we first show that
\wonerma is \NP-hard for graphs consisting only of a perfect matching by a reduction from
\onerma.
The hardness of \wonerma on $\mathcal{K}^*$ and $\mathcal{P}^*$ follows from this result.

Before we give the proof of Theorem~\ref{thm:weighted-main}, we need the
following key lemma. The polynomial-time algorithm described in the proof of
the lemma uses the fact that \dsnp can be solved in polynomial
time if the number of terminal pairs is constant~\cite{feldman_ruhl_06}.

\begin{lemma}
	\label{lemma:weighted-trees}
	Let $r \in \nat$ be constant and let $\mathcal T$ be a class of perfectly matchable trees, each with at most $r$ leaves. Then \wonerma on $\mathcal T$ admits a
	polynomial-time algorithm.
\end{lemma}

We remark that the running time of the algorithm given in
Lemma~\ref{lemma:weighted-trees} slicewise polynomial in the number of leaves of the input graph.
We can now state the proof of our main result.

\begin{proof}[Proof of Theorem~\ref{thm:weighted-main}]
	According to Lemma~\ref{lemma:hardness:stars-and-paths}, \wonerma is NP-hard if $\mathcal G$ completely contains the class $\mathcal{K} = \{K_{1,r}^* \mid r \in \mathbb N\}$ or the class $\mathcal{P} = \{P_r^* \mid r \in \mathbb N\}$. Assuming $\P\neq\NP$, this proves the \emph{only if} statement of the theorem.
	
	To see the \emph{if} statement, let us consider $r \in \mathbb N$ such that $\mathcal G$ does not contain $K_{1,r}^*$ or $P_r^*$.
	First we will reduce the problem to the case when $\mathcal G$ contains only trees.
	For this, let $\mathcal T$ be the class of all trees in $\mathcal G$ that admit a perfect matching.
	
        \setcounter{claim}{0}
        \begin{claim}
            There is a polynomial time reduction of \wonerma on $\mathcal G$ to \wonerma on $\mathcal T$.
            \label{clm:reductiom-to-trees}
        \end{claim}
	The key idea for the proof is to define an equivalent instance on an arbitrary tree of $G$ on 
	an adapted cost function.
We may hence restrict our attention to \wonerma on the class $\mathcal T$.
	As the next claim shows, the relevant trees contained in $\mathcal T$ have a bounded number of leaves.
	
        \begin{claim}\label{clm:leaf-number}
            There is some number $f(r)$ depending only on $r$ such that every tree in $\mathcal T$ has at most $f(r)$ many leaves.
        \end{claim}
        According to the above claims, there is a polynomial reduction of \wonerma on $\mathcal G$ to \wonerma on a class of trees with a bounded number of leaves.
        Hence, Lemma~\ref{lemma:weighted-trees} implies that \wonerma on $\mathcal G$ can be solved in polynomial time.
\end{proof}

\section{Conclusion}
\label{sec:conclusion}
We presented algorithms for the task of securing matchings
of a graph against the failure of a single edge. For this, we established a connection
to the classical strong connectivity augmentation problem.  
Not surprisingly, the unit weight case is more accessible, and we were able to
give a $\log_2 n$-factor approximation algorithm, as well as polynomial-time
algorithms for graphs of bounded treewidth and chordal-bipartite graphs.
For general non-negative weights, we showed a close relation to \dsnp in terms of approximability
and gave a dichotomy theorem characterizing
minor-closed graph classes which allow a polynomial-time algorithm.

In our opinion, the case of a single edge failure is well
understood now and so one might go for the case of a constant number of edge failures next.
Let us remark that if the number of edge failures is a part of the input, even checking feasibility is \NP-hard~\cite{dourado_et_al_15,LMMP:12}.

\bibliography{references}

\appendix

\newpage
\section{Omitted Proofs}
\subsection{Proofs Omitted from Section \ref{sec:connectivity}}

\begin{proof}[Proof of Fact~\ref{fact:iso2}]
  Let $M$ and $M'$ be two distinct perfect matchings of $G$. Then their
  symmetric difference $M \operatorname{\Delta} M'$ is a sum of $(M, M')$-alternating cycles.
  But each cycle is in some strong component of $D(G, M)$ and $D(G, M')$, so
  both condensations must be isomorphic.
\end{proof}

\begin{proof}[Proof of Fact \ref{fact:sourcesink}]
  Let $vw$ be an arc in $L$. Let $L'$ be a copy of $L$, where the arc $vw$ is
  replaced by an arc $v'w'$ from a sink $v'$ of $C(D(G, M))$ reachable from $v$
  to a source $w'$ of $C(D(G, M))$ from which $w$ is reachable. We show that $G + L'$ is
  robust. Suppose for a contradiction that this is not the case. Then there is
  some edge $xy \in M$, such that $x \in U$, $y \in W$, and $xy$ is not on
  an $M$-alternating cycle in $G+L'$. Equivalently, $x$ is not contained in a
  directed cycle of~$D+L'$.  However, since $G+L$ is robust, we have that $x$ and the arc $vw$ are contained in 
  some directed cycle $C = v_1, v_2, \ldots, v_k = v_1$ of $D+L$. That is,
  there are $1 \leq i, j < k$, such that $x = v_i$, $v = v_j$, and $w =
  v_{j+1}$. Let $P$ ($Q$) be a path connecting $v$ and $v'$ ($w'$ and $w$).
  Then $C' := v_1, v_2,\ldots, v_{j-1}, P, Q, v_{j+2}, \dots, v_k$ is a
  closed walk that contains a simple directed cycle visiting $x$.  This
  contradicts our assumption that $x$ is not on a directed cycle in $G+L'$. By
  iterating this argument we obtain an arc-set $L'$ such that $|L'| \leq |L|$ and
  $G+L'$ is robust. By construction, $L'$ contains only arcs that
  connect sources and
  sinks of $C(D(G, M))$.
\end{proof}

\begin{proof}[Proof of Fact \ref{fact:eswaran-tarjan-and-optimal-solutions}]
    By assumption, we have that $C(D(G, M))$ contains no strong sources or
    sinks. Therefore, each source and each sink of $C(D(G, M))$ corresponds to
    a \unsafe edge of the matching $M$. Let $L' \subseteq \overline{E}$ of
    minimum cardinality, such that $G + L'$ is robust. By
    Fact~\ref{fact:sourcesink}, we may assume that $L'$ connects only sinks to
    sources of $C(D(G, M))$. If~$|L'| < \gamma(D(G, M)) = \max \{
    |\sources(C(D(G, M)))|, |\sinks(C(D(G, M)))|\}$, then at least one sink or
    at least one sources is not incident to an arc of $L'$. Therefore, the graph $G + L'$ is not robust. 
\end{proof}

\subsection{Proofs Omitted from Section \ref{sec:sourcecover}}
\begin{proof}[Proof of Theorem~\ref{thm:sc:chordal-bipartite}]
  Let $(D)$ be a \sourcecover instance such that $\undirected(D)$ is connected,
  has at least one arc,  and $\undirected(D)$ contains no induced cycle of
  length at least six. If $\undirected(F(D))$ is chordal-bipartite, then we can
  apply the polynomial-time algorithm for \setcover on chordal-bipartite
  incidence graphs, see~\cite[pp.~562-573]{nemhauser_wolsey} and~\cite{HKS:85}. It
  remains to show that
  $\undirected(F(D))$ is chordal-bipartite.  Suppose for a contradiction, that
  $\undirected(F(D))$ contains an induced cycle $C_{FD} = \{ s_1, t_1, \ldots,
  s_k, t_k, s_{k+1}= s_1 \}$, where $s_1, s_2, \ldots, s_k \in \sources(F(D))$
  and $t_1, t_2, \ldots, t_k \in \sinks(F(D))$, and $k \geq 3$.  In order to
  keep the notation concise, let $t_0 := t_k$.

  Since $C_{FD}$ is a cycle in $\undirected(F(D))$ connecting sources and
  sinks, we have that for $1 \leq i \leq k$, there are directed paths
  $P_i^{i-1}$ and $P_i^{i}$ in $D$ such that $P_i^{i-1}$ connects $s_i$ to
  $t_{i-1}$ and $P_i^{i}$ connects $s_i$ to $t_{i}$. We now construct a cycle
  $C$ in $\undirected(D)$ and then show that $C$ is chordless and has length at
  least $k$. Let $Q_1^1$ be any shortest path from $s_1$ to $t_1$ in $D$. Let
  us assume we already constructed the paths $Q_j^j$ and $Q_j^{j-1}$ for $1
  \leq j \leq i \leq k-1$.  We now define the paths $Q_{i+1}^{i}$ and
  $Q_{i+1}^{i+1}$ in the following way: $Q_{i+1}^{i}$ is a shortest path from
  $s_{i+1}$ to $Q_i^i$ in $D$. If there exist more than one shortest path, then
  we pick the one whose endpoint is closest to $y_{i}$ on $Q_i^i$. We refer to
  this endpoint by $t_i'$.  Similarly, $Q_{i+1}^{i+1}$ is a shortest path from
  $Q_{i+1}^{i}$ to $t_{i+1}$ in $D$. If there is more than one shortest path,
  then we pick the one whose starting point is closest to $t_{i}$ on $Q_i^i$.
  We refer to this starting point by $s_i'$.  Finally $Q_1^k$ (= $Q_1^0$) is a
  shortest path from $Q_1^1$ to $Q_k^k$.  Again, if there is more than one such
  shortest path, then we first pick the one whose starting point is closest to
  $t_1$ on $Q_1^1$ and then whose endpoint is closest to $t_k$ on $Q_k^k$.  We
  refer to these two vertices by $s_1'$ and $t_k'$, respectively. Now let $C =
  \{ Q_1^1, Q_2^1, \ldots, Q_k^{k-1}, Q_k^k, Q_1^k \}$. 
  
  We have that $C$ is by construction a cycle in $\undirected(D)$.  Note that
  $s_i' \neq t_{i-1}'$ and $s_i' \neq t_{i}'$, since otherwise $s_{i-1}$ were
  adjacent to $t_i$ or $s_{i+1}$ were adjacent to $t_{i-1}$ in $\undirected(F(D))$.
  Therefore, $C$ is simple and has length at least $k$. Now assume for a
  contradiction that $C$ has some chord $a$. Observe that $a$ connects two
  distinct paths $Q_i^j$ and $Q_k^l$ (without loss of generality, $i \leq k$
  and $j \leq \ell$) only if $i = k$ and $j = \ell -1$ or $i = k-1$ and $j =
  \ell$, respectively, since otherwise $C_{FD}$ is not chordless. On the other
  hand $i = k$ and $j = \ell-1$  contradicts the
  choice of the starting vertex of $Q_i^i$ on $Q_i^{i-1}$.  Similarly, $i =
  k-1$ and $j = \ell$ contradicts the choice of the endvertex of $Q_{i+1}^i$ on
  $Q_i^i$. Therefore, $C$ is an induced cycle in $\undirected(D)$ of length at
  least $k$, which contradicts our assumption that $\undirected(D)$ has no
  induced cycles of length $\geq 6$.
\end{proof}

\subsection{\sourcecover on graphs with bounded treewidth}

We now present the \FPT-time algorithm for \nwdstp on acyclic digraphs that is
single-exponential in the treewidth of the underlying undirected graph and
linear in the instance size. Let us first again recall some
definitions. The problem node-weighted \dstp problem is defined as follows.

\begin{quote}
  \textsc{\nwdstp}\\
  \textbf{instance:} Acyclic digraph $D = (V, A)$, costs $c \in \mathbb{R}_{\geq 0}^V$, terminals $T \subseteq V$, root $r \in V$.\\
  \textbf{task:} Find a minimum-cost subset $F \subseteq V$, such that $r$ is connected to each terminal in $(F, E(F))$.
\end{quote}

Our algorithm is presented best using a so-called \emph{nice tree decomposition}.
This kind of decomposition limits the structure of the difference of two adjacent nodes in the decomposition.
Formally, consider a tree decomposition $T$ of a graph $G$, rooted in a leaf of $T$.
We say that $T$ is a \emph{nice tree decomposition} if every node $x \in V(T)$ is of one of the following types.
\begin{itemize}
    \item \textbf{Leaf:} $x$ has no children and $B_x=\emptyset$.
    \item \textbf{Introduce:} $x$ has exactly one child $y$ and there is a vertex $v \notin B_y$ of $G$ with $B_x = B_y \cup \{v\}$.
    \item \textbf{Forget:} $x$ has exactly one child $y$ and there is a vertex $v \notin B_x$ of $G$ with $B_y = B_x \cup \{v\}$.
    \item \textbf{Join:} $x$ has two children $y$ and $z$ such that $B_x=B_y=B_z$.
\end{itemize}
Such a nice decomposition is easily computed given any tree decomposition of $G$.
We define $x^+$ to be the subtree of $T$ rooted in $x$: the tree of all vertices not connected to the root in the forest $T-x$, together with $x$.
By $B_x^+$ we denote the set of vertices contained in all bags of nodes in $x^+$.

A \emph{coloring} of a bag $B_x$ is a mapping $f: B_x \rightarrow \{ 1, 1_?, 0 \}^{\vert B_x \vert }$, where the individual colors have the following meaning.
\begin{itemize}
\item Active and already covered, represented by a 1, means that the vertex is active
and that there is at least one predecessor of it that is either labeled 1 or $1_?$.
\item Active and not yet covered, represented by a $1_?$, means that the vertex is active
but every predecessor is labeled 0.
\item Not active, represented by a $0$, means that the vertex is not contained in the solution.
\end{itemize}

Note that there are $3^{\vert B_x \vert}$ colorings of the bag $B_x$.
For a coloring $f$ of $x$ we denote by $\mbox{OPT}(f,x)$ the minimum cost\footnote{Here, a vertex $v$ has a cost $c(v)$ if it is colored $1$ or $1_?$ and 0 otherwise.} of a coloring $B_x^+ \to \{ 1, 1_?, 0 \}$ satisfying the following conditions.
\begin{enumerate}[(a)]
	\label{tw:conditions-a-d}
    \item each vertex in $B_x$ is colored 1, $1_?$ or 0 according to $f$.
    \item every vertex of $B_x^+ \setminus B_x$ is colored 0 or 1.
    \item each sink $v \in V^- \cap B_x^+$ is colored either 1 or $1_?$.
    \item each $v \in B_x$ with $f(v) = 1$ is either a source or at least one predecessor of $v$ in $D(B_x^+)$ is colored
    either $1$ or $1_?$.
\end{enumerate}

To present the individual steps of the algorithm, assume that we are given a nice tree decomposition of our input graph.
Let us say we are currently considering the node $x$ in $T$
and distinguish between the type of node $x$.
\begin{itemize}
    \item \textbf{Leaf:} put $\mbox{OPT}(f,x) = 0$ if it is not the root.
    \item \textbf{Introduce:} let $y$ be the unique child of $x$ and let $v \notin B_y$ such that $B_x = B_y \cup \{v\}$.
    The value $\mbox{OPT}(f,x)$ depends on the type of vertex $v$ is and on the coloring $g$ of $y$.
    By definition, sinks have to be active and therefore the optimal value is $\infty$ if $f(v)=0$.
    The same is true for sources labeled $1_?$ in $f$ (those do not have predecessors and need to be labeled either 1 or 0).
    Finally, we set the cost to be $\infty$ if $v$ is labeled 1 in $f$ and not a source, but non of its predecessors is
    active in $f$.
    Thus we set
    \begin{equation}\label{eqn:introduce-node}
\mbox{OPT}(f,x) =
\begin{cases}
\infty, \text{ if } v \in V^- \text{ and } f(v) = 0\\
\infty, \text{ if } v \in V^+ \text{ and } f(v) = 1_?\\
\infty, \text{ if } v \notin V^+ \text{ and } f(v) = 1 \text{ and } (\delta^-(v) \cap B_y) \subseteq f^{-1}(0)\\
\min\{ \mbox{OPT}(g, x) : (g, y) \text{ is compatible to } (f, x) \}, \text{ if } f(v) = 0\\
\min\{ \mbox{OPT}(g, x) : (g, y) \text{ is compatible to } (f, x) \} + c(v), \text{ else,}
\end{cases}
\end{equation}
where the pair $(g, y)$ is compatible to $(f, x)$ if the following conditions hold.
\begin{itemize}
\item If $f(v) = 0$, then $g = f|_{B_y}$.
As the introduced vertex is not considered to be part of the solution,
we can simply keep the coloring of the child node.
\item If $f(v) = 1_?$, then $f^{-1}(0) = g^{-1}(0)$, $f^{-1}(1) = g^{-1}(1) \cup (g^{-1}(1_?) \cap \delta^+ (v))$, and $\delta^-(v) \subseteq g^{-1}(0)$.
This condition makes sure that the introduced vertex can only be labeled $1_?$ if none of its predecessors is labeled 1 or $1_?$.
\item If $f(v) = 1$, then $f^{-1}(0) = g^{-1}(0)$, $f^{-1}(1) = g^{-1}(1) \cup (g^{-1}(1_?) \cap \delta^+ (v))$, and, moreover, $\delta^-(v) \setminus g^{-1}(0) \neq \emptyset$ or $v \in V^+$.
This conditions says that the introduced vertex can only be labeled $1$ if at least one of its predecessors is labeled 1 or $1_?$, unless it is a source.
\end{itemize}
    \item \textbf{Forget:} let $y$ be the unique child of $x$ and let $v \notin B_x$ such that $B_y = B_x \cup \{v\}$.
    Then we put
\begin{equation*}
\label{eqn:forget-node}
    \mbox{OPT}(f,x) =
\min \{ \mbox{OPT}(g, y) : f = g|_{B_x} \} \text{ if } g(v) \neq 1_?.
\end{equation*}
We do not allow a vertex labeled $1_?$ to be forgotten, as we can not assure to cover it in later bags.
For the remaining cases we simply keep the optimal value.
    \item \textbf{Join:} let $y$ and $z$ be the two children of the join node $x$ with $B_x = B_y = B_z$.
    We put
\begin{equation}\label{eqn:join-node}
    \mbox{OPT}(f,x) =
\min \{ \mbox{OPT}(g, y) + \mbox{OPT}(h, z) - \sum_{v \in B_X \setminus (g^{-1}(0) \cap h^{-1}(0))} c(v)\},
\end{equation}
where the minimum runs over all colorings $g$ of $y$ and $h$ of $z$ with
$f^{-1}(0) = g^{-1}(0) = h^{-1}(0)$ and $f^{-1}(1) = g^{-1}(1) \cup h^{-1}(1)$.
    \item \textbf{Root:} as the graph is connected and the root node is a leaf, the root node is a forget node,
    where its child node contains exactly one vertex in its bag.
     The algorithm terminates with the output
    \[
    \mbox{OPT} = \mbox{OPT}(f, x),
    \]
\end{itemize}
where $f$ is the unique coloring of the empty bag $x$.

Having presented the algorithm, we need to prove Theorem \ref{thm:sc:tw} by 
showing the correctness and bounding the running time of the algorithm.

\begin{proof}[Proof of Theorem \ref{thm:sc:tw}]
We need to show that the algorithm works correctly and is fixed parameter tractable when parameterized by the treewidth of the underlying graph.
Let $T$ be a nice tree decomposition of $\undirected(D)$ of width $w$ with $t$ nodes.
\setcounter{claim}{0}
\begin{claim}
    The algorithm correctly computes an optimal solution to \textsc{Node Weighted Directed Steiner Tree in Acyclic Graphs}.
    \label{clm:tw correctness}
\end{claim}
\begin{proof}
    We show the statement by a straight-forward inductive proof on the decomposition tree.
    The induction hypothesis states that $\mbox{OPT}(f,x)$ is the minimum cost of a solution
    induced by the vertices of $B_x^+$, satisfying the conditions \textbf{(a)}-\textbf{(d)} (see~p.~\pageref{tw:conditions-a-d}).
    The base case are the leaf nodes where the hypothesis clearly holds.
    Now let the induction hypothesis be true for all descendants of $x$.
    We distinguish between the remaining three node types and argue that the induction hypothesis holds in $x$.
    \begin{itemize}
        \item \textbf{Introduce:} let $y$ be the unique child of the introduce node $x$
            and let $v \notin B_y$ such that $B_x = B_y \cup \{v\}$.
            Clearly \textbf{(a)} holds and \textbf{(b)} holds by the induction hypothesis.
            By putting $\mbox{OPT}(f,x)$ to $\infty$ if $f(v) = 0$ for a sink $v \in V^-$, \textbf{(c)} also holds.

            For \textbf{(d)} observe that the notion of compatibility is defined correctly. If $f(v) \in \{1_?, 0 \}$ this is trivial.
            For $f(v)= 1$ observe that $v$ has to satisfy the condition that $\delta^-(v) \setminus g^{-1}(0) \neq \emptyset$.
            Thus the condition \textbf{(d)} holds for $x$.
            Now for a given coloring $f$ we have to check if $\mbox{OPT}(f,x)$ is calculated correctly.
            This is true for the cases in which $\mbox{OPT}(f,x)$ is set to $\infty$. So it remains to
            show that we identify all compatible colorings $g$ for $y$ to calculate the minimum.
            The case $f(v) = 0$ is trivial.
            For the cases $f(v) \in \{ 1, 1_? \}$ observe that $g$ has to satisfy
            $f^{-1}(0) = g^{-1}(0)$ and $f^{-1}(1) = g^{-1}(1) \cup (g^{-1}(1_?) \cap \delta^+ (v))$.
            Calculating the minimum over all pairs $(g, y)$ compatible to $(f, x)$ is hence correct.
            Finally it is clear that $\mbox{OPT}(f,y)$ that we have to add $c(v)$ to the minimum of all compatible colorings
            $(g, y)$ for $(f, x)$ if $f(v) \neq 0$.
        \item \textbf{Forget:} let $y$ be the unique child of $x$ and let $v \notin B_x$ such that $B_y = B_x \cup \{v\}$.
            For a forget node we put $\mbox{OPT}(f,x) = \min \{ \mbox{OPT}(g, y) : f = g|_{B_x} \} \text{ if } g(v) \neq 1_?$.
            Clearly \textbf{(a)}, \textbf{(c)} and \textbf{(d)} hold by the induction hypothesis.
            \textbf{(b)} also holds as we only allow colorings that satisfy $f(v) \neq 1_?$.
            Finally it is easily verified that the calculation of $\mbox{OPT}(f,x)$ is correct.
        \item \textbf{Join:} let $y$ and $z$ be the two children of the join node $x$ with $B_x = B_y = B_z$.
            By \eqref{eqn:join-node}, a vertex $v \in B_x$ may only be colored $1$ if it is colored 1
            either in $B_y$ or $B_z$. As the induction hypothesis holds for $y$ and $z$, \textbf{(a)}-\textbf{(d)}
            also hold for $x$. It remains to show that $\mbox{OPT}(f,x)$ is calculated correctly.
            The considered colorings $g$ and $h$ of $y$ and $z$ have to satisfy
            $f^{-1}(0) = g^{-1}(0) = h^{-1}(0) \text{ and } f^{-1}(1) = g^{-1}(1) \cup h^{-1}(1)$.
            By adding $\mbox{OPT}(g, y) + \mbox{OPT}(h, z)$ we count the vertices 
            in the set $B_X \setminus (g^{-1}(0) \cap h^{-1}(0))$
            twice. Thus we obtain
            $\mbox{OPT}(f,x) = \mbox{OPT}(g, y) + \mbox{OPT}(h, z) - \sum_{v \in B_X \setminus (g^{-1}(0) \cap h^{-1}(0))} c(v)$.
    \end{itemize}
\end{proof}

\begin{claim}
    Given $T$, the running time of the dynamic programming algorithm is bounded by $O(5^{w}t)$.
    \label{clm:tw-running-time}
\end{claim}
\begin{proof}
    In each node $x$ of the nice tree decomposition $T$ we consider $O(3^{|B_x|})$ many different colorings $f$.
    We bound the running time for a bag by considering the different kinds of bags.
    For this, note that the interesting steps are the computation of the pairs $(g,y)$ compatible to $(f, x)$
    for the minimum in \eqref{eqn:introduce-node},
    and the computation of the minimum in \eqref{eqn:join-node}.

    Consider an introduce node $x$ with its unique child $y$ and let $v \notin B_y$ such that $B_x = B_y \cup \{v\}$.
    Let $f$ and $g$ be colorings for $x$ and $y$, respectively. For a vertex $u \in B_y$ we consider all possible
    combinations $(f(u), g(u))$ for the three possible values of $f(v)$ which are given by \eqref{eqn:introduce-node}.
    \begin{itemize}
        \item In the case $f(v) = 0$ we have that $g = f|B_x$, that is, $(f(u), g(u)) = (g(u), g(u))$.
        \item In the case $f(v) = 1_?$ we have that
            \begin{equation}
                (f(u), g(u)) \in
                \begin{cases}
                    \{(0, 0),(1, 1),(1, 1_?)\}, & \text{ if } u \in \delta^+(v),\\
                    \{(0, 0),(1, 1),(1_?, 1_?)\}, & \text{ if } u \notin \delta^+(v),
                \end{cases}
            \end{equation}
            and $(f,x)$ and $(g,y)$ are not compatible unless $\delta^-(v) \subseteq g^{-1}(0)$.
        \item In the case $f(v) = 1$ we allow the same pairs $(f(u), g(u))$ like in the case $f(v) = 1_?$, but $(f,x)$ and $(g,y)$ are not compatible if
            $\delta^-(v) \subseteq g^{-1}(0)$.
    \end{itemize}
    We basically have three different options for the pairs $(f, g)$. Processing through $f$ and $g$ at the same time
    leads to the total running time for an introduce node of at most $O(3^w)$.

    For a join node, let $y$ and $z$ be the two children of $x$ with $B_x = B_y = B_z$.
    Let $f, g, h$ be colorings of $x, y$ and $z$, respectively. For a vertex $u \in B_x$ we consider all possible
    combinations $(f(u), g(u), h(u))$ with
    \begin{equation}
        (f(u), g(u), h(u)) \in \{ (0, 0, 0), (1_?, 1_?, 1_?), (1, 1_?, 1), (1, 1, 1_?), (1, 1, 1) \}.
    \end{equation}
    Here we are given five different options for the triples $(f, g, h)$, and so the total computation time is at most $O(5^w)$.

    The overall bottleneck case is when $x$ is a join-node since we need to compute \eqref{eqn:join-node}.
    As we just said, this can be done in $O(5^{w})$ time. Since we have $t$ nodes, the total running time
    is $O(5^{w} t)$.
\end{proof}

By storing the best current solution alongside the $\mbox{OPT}(f,x)$-values 
we can compute an optimal solution together with \OPT.
We do not give details here since this is standard.
Finally observe that the algorithm is indeed fixed parameter tractable when 
parameterized by the treewidth of the underlying graph.
This completes the proof. 
\end{proof}

Given a graph on $n$ vertices of treewidth $w$, one can compute a tree
decomposition of width $w$ in time $O(2^{O(w^3)}n)$ by Bodlaender's famous
theorem~\cite{DBLP:journals/siamcomp/Bodlaender96}. Given a tree decomposition
of width $w$ with $t$ nodes, one can compute a nice tree decomposition of width
$w$ on $O(wt)$ nodes in $O(w^2t)$ time in a straightforward way.  We thus
arrive at an algorithm that, given a tree decomposition of width $w$, runs in
$O(5^w w |V|)$ time.

\subsection{Proofs Omitted from Section \ref{sec:onerma}}

\begin{proof}[Proof of Corollary~\ref{cor:onerma:approx}]
  Let $\I = (G, M)$ be an instance of \onerma. We use
  Theorem~\ref{lemma:main} to obtain from $\I$
  in polynomial time the \sourcecover instances $\mathcal{A}_1$ and
  $\mathcal{A}_2$ such that $\OPT(\I) = \OPT(\I') = \max \{\mathcal{A}_1,
  \mathcal{A}_2\}$. Let $i \in \{1, 2\}$. Let $S_i$ be the acyclic input graph of
  $\mathcal{A}_i$. We ``flatten'' the graph $S_i$  as described in
  Section~\ref{sec:sourcecover} to obtain a \setcover instance $\mathcal{B}_i$
  on the incidence graph $\undirected(F(S_i))$. The classical greedy algorithm
  for \setcover yields $((\ln |M|) + 1)$-approximate cover $C_i$ for
  $\mathcal{B}_1$.  By Theorem~\ref{lemma:main}, we can construct from $C_1$ and
  $C_2$ in polynomial time a solution $L$ of $\I$.  By recalling that $n =
  |V(G)| \geq |M|/2$ and some simple calculations, we conclude that $L$ is
  $\log_2 n$-approximate.
\end{proof}

\begin{proof}[Proof of Corollary \ref{cor:onerma:tw-cb}]
  Let $k \in \nat$ and $\I = (G, M)$ be an instance of \onerma such that the
  graph $G = (U + W, E)$ has treewidth at most $k$.  We then use
  Theorem~\ref{lemma:main}, to construct in polynomial time the \sourcecover
  instances $\mathcal{A}_1 = (S_1)$ and $\mathcal{A}_2 = (S_2)$ from $\I'$. By
  Theorem~\ref{lemma:main}, $\undirected(S_1)$ and $\undirected(S_2)$ are minors
  of $G$. Since treewidth is monotone under taking minors, we have that
  $\undirected(S_1)$ and $\undirected(S_2)$ have treewidth at most $k$.  Hence,
  by Theorem~\ref{thm:sc:tw}, optimal solutions of $\mathcal{A}_1$ and
  $\mathcal{A}_2$ can be computed in polynomial time. By
  Theorem~\ref{lemma:main}, we can obtain in polynomial time from these two
  solutions a solution $L$ of \I, such that $|L| = \OPT(\I) = \max \{
  \OPT(\mathcal{A}_1), \OPT(\mathcal{A}_2) \}$.

  Now let $\I = (G, M)$ be an instance of \onerma such that $G$ is
  chordal-bipartite.  Then $\undirected(D(G,M))$ contains no induced cycle of
  length at least six.  To see this, note that this is a special case in the
  proof of Claim~\ref{claim:noc6} in the proof of Proposition
  \ref{thm:onerma:reduction}.  We use Theorem~\ref{lemma:main}, to construct in
  polynomial time the \sourcecover instances $\mathcal{A}_1 = (S_1)$ and
  $\mathcal{A}_2 = (S_2)$ from \I.  In order to obtain the source cover
  instances, we simply contract all edges of a strong component of $D(G, M)$ to
  a single vertex. As the contraction of edges only reduces the size of cycles,
  the underlying undirected graphs occurring in the source cover instances
  cannot have induced cycles of length at least six.  Hence, by Theorem
  \ref{thm:sc:chordal-bipartite}, optimal solutions of $\mathcal{A}_1$ and
  $\mathcal{A}_2$ can be computed in polynomial time. By
  Theorem~\ref{lemma:main}, we can obtain in polynomial time from these two
  solutions a solution $L$ of \I, such that $|L| = \OPT(\I) = \max \{
  \OPT(\mathcal{A}_1), \OPT(\mathcal{A}_2) \}$.
\end{proof}

\begin{proof}[Proof of Proposition \ref{thm:onerma:reduction}]
    Let $M$ be a maximum matching of $G = (U + W, E)$.  Without loss of
    generality, we assume that $M$ is $U$-perfect, so $|U| \leq |W|$.
    Otherwise, adding an edge joining two unmatched vertices solves the
    problem. We construct the graph $G' = f(G)$ as follows. Let $G'$ be a copy
    of $G$ to which we add a leaf to each unmatched vertex of $W$. We then add
    a vertex $z$ to $U$ joined to each vertex of the other part of the
    bipartition. Finally, we add a vertex $z'$ joined to $z$ and each leaf
    added in the previous step. Furthermore, we extend the matching $M$ of $G$
    to a perfect matching $M'$ of $G'$ by adding the edges between the leaves
    and the previously unmatched vertices to $M'$.  Note that by construction,
    if $e$ is a \unsafe edge of $G'$ then $G - e$ does not admit a matching of
    cardinality $|M|$. 

  We prove the statements one by one.
  \setcounter{claim}{0}
  \begin{claim}
      $\OPT(\I') = \OPT(\I)$ and from a solution $L'$ of $\I'$
      we can construct in polynomial time a solution $L$ of $\I$ such that $|L| \leq |L'|$.
  \end{claim}
  \begin{proof}
      Let $(U, W)$ be the bipartition of $G$ as chosen in the construction, i.e., such that $z \in U$. Note
      that since $z$ is joined to each vertex $w \in W$, there is an arc from each
      vertex of $D(G', M')$ to $z$. Therefore, $C(D(G', M'))$ has a single
      strong sink, say $\widehat S$, originated from the vertex set $\widehat Y \subseteq V(D(G', M'))$.
      Observe that $z, u_1', u_2', \ldots, u_k' \in \widehat Y$. For a strong
      component $s$ of $D(G',M')$, let $Y_s$ be the set of vertices of $V(D(G', M'))$ in
      the component $s$.

      We first show that $\OPT(\I) \leq \OPT(\I')$. Let $\tilde L$ be a solution of
      $\I'$. According to Fact~\ref{fact:sourcesink} and the algorithm contained in
      its proof we may construct from $\tilde L$ a solution $L'$ to $\I'$ of
      cardinality at most $|\tilde L|$, such that $L'$ connects only sources and
      sinks of $C(D(G', M'))$. Since there is only the sink $\widehat S$, we may further
      assume that $L'$ connects $\widehat S$ to a selection $S \subseteq \sources(D(G',
      M'))$. Let $x \in W$ be $M$-exposed. We construct a solution $L$ of \I as
      follows. For each source $s \in S$, we pick a vertex $u \in U$ in the
      corresponding component in $D(G', M')$ and add the edge $ux$ to $L$.  We
      now show that $G+L$ is robust. Recap that by construction, the
      \unsafe edges of $(G', M')$ are precisely the \unsafe edges of $(G,
      M)$.  Let $e \in M$ be a \unsafe edge of $(G, M)$.  Since $L'$ is feasible
      for $\I'$, any vertex $u \in U$ that is incident to a \unsafe edge of
      $(G', M')$ is reachable from some $s \in S$ by a directed path in $C(D(G',
      M'))$. This directed path corresponds to an $M$-alternating path in $G$
      starting from any vertex $u \in Y_s$ with an $M$-edge. Therefore, the edge
      $e$ is not \unsafe in $(G + ux, M)$ for any $u \in Y_s$. Hence, $(G + L,
      M)$ has no \unsafe edges and from $|L| = |L'| \leq |\tilde L|$ we conclude
      that $\OPT(\I) \leq \OPT(\I')$. Moreover, we can construct $L$ from $L'$ in
      polynomial time.

      It remains to show that $\OPT(\I') \leq \OPT(\I)$. Let $L$ be an optimal
      solution of \I. Note that each \unsafe edge of $(G, M)$ is on an
      $M$-alternating cycle or a maximal even-length $M$-alternating path in $G +
      L$. We construct from $L$ a solution $L'$ to $\I'$. Let $x \in W$ be
      $M$-exposed. For each $u \in U$ and $w \in W$ such that $uw \in L$, we add
      the edge $ux$ to $L'$. We show that $L'$ is feasible for $\I'$. Let $uw \in
      L$ and let $e \in M$ be a \unsafe edge of $(G, M)$ on a maximal
      $M$-alternating path $P$ of even length. By replacing $uw$ by $ux$, we split
      $P$ into at most two maximal $M$-alternating paths of even length. Oh the
      other hand, suppose that $e$ is on some $M$-alternating cycle involving $uw$.
      Replacing $uw$ by $ux$ yields a maximal $M$-alternating path containing $e$.
      Therefore, each \unsafe edge of $(G, M)$ is on some maximal
      $M$-alternating path of even length in $G + L'$.  By the construction above,
      each \unsafe edge of $(G', M')$ is hence on some maximal $M'$-alternating
      cycle of $(G' + L', M')$, so $M'$ is robust in $G' + L'$. Since $|L| =
      |L'|$, we have that $\OPT(\I) \leq \OPT(\I')$.
  \end{proof}

  \begin{claim}
  $\tw(G') \leq \tw(G) + 2$.
  \end{claim}  
  \begin{proof}
      To prove Claim~\ref{itm:onerma:tw}, observe that adding a single vertex
      to a graph increases its treewidth by at most one. Furthermore, adding a leaf
      vertex to a graph does not increase its treewidth. We obtain $G'$ from $G$ by
      adding leaf vertices to each exposed vertex and finally add two more
      vertices.  Therefore, $\tw(G') \leq \tw(G) + 2$.
  \end{proof} 

  \begin{claim}
  If $G$ is chordal-bipartite then $\undirected(D(G', M'))$ has no induced cycle of length at least six.
  \label{claim:noc6}
  \end{claim}   
  \begin{proof}
      Now suppose that $G$ is chordal-bipartite. Assume for a contradiction that
      $H = \undirected(D(G', M'))$ has an induced cycle $C'$ of length at least six.
      It is easy to see that $z$ is not contained in $C'$ since $z$ is adjacent to all $v \in H$.
      In order to obtain a cycle $C$ in $G$, for every edge $e$ in $H[C']$, replace $e$ by the unique
      corresponding path $P_e$ in $G'$ consisting of a matching edge and a non-matching edge.
      If two consecutive paths $P_e$ and $P_{e'}$ use the same matching edge,
      simply delete those matching edges in $C$ such that $C$ is a cycle.
      Note that all edges in $H[C']$ incident to $U'$
      are directed from $U'$ to $U$ in $D(G', M')$. Hence consecutive edges to vertices in $U'$
      use the same matching edges, which are then deleted. Therefore $V(C) \subseteq V(G)$.
      Now if $G[C]$ contains a chord then $H[C']$ also contains a chord due to fact \ref{fact:arc-edge-correspondence}.
      Therefore $C$ is an induced cycle in $G$ (since $z \notin C'$) and $|C|\geq|C'|\geq 6$, a contradiction.
  \end{proof}
\end{proof}

\subsection{Proofs Omitted from Section \ref{sec:wonerpma}}

\begin{proof}[Proof of Proposition~\ref{prop:hardness:onerma-dsf}]
We first prove the following statement: An $f(n')$-factor approximation algorithm for \wonerma 
implies an $f(n+k)$-factor approximation algorithm for \dsnp.
Let \I be a feasible instance of \dsnp with input graph $D=(V, A)$, $\vert V \vert = n$, costs $c
\in \mathbb{Z}^A_{\geq 0}$ and $k$ terminal pairs $(s_1, t_1), \ldots, (s_k, t_k) \in
V$.  Without loss of generality, let $S= \bigcup_{i
\in [k]} \{s_i\}$ be the set of sources and let $T= \bigcup_{i \in [k]} \{t_i\}$ be the set
of sinks of $D$.  
We may also assume that $(s_i, t_i) \notin A$ for all $i \in [k]$.
In the reduction it is important that each terminal is a unique vertex, i.e.
$t_i \neq t_j$ for all $i \neq j$, $i, j \in [k]$.
We ensure this by introducing a copy of each terminal $t_i$ and then connect it 
to all neighbors of the original vertex, resulting in a graph of at most $n+k$ vertices.

To obtain an instance $\I'$ of \wonerma, we create a bipartite graph
$G=(U + W, E)$, a cost function $c' \in \mathbb{Z}^{\overline{E}}_{\geq 0}$, and
a perfect matching $M$ of $G$ in the following way.

For each $v \in V$ we add the vertices $u_v$ and $w_v$, and the edge $u_v w_v$ to $G$ and $M$. 
For each $i \in [k]$ we additionally add the edge $u_{s_i} w_{t_i}$ to $E$.
For each matching edge $u_v w_v \in M$ with $v \notin \{t_1, \ldots, t_k\}$ we add the vertices $u_v'$ and $w_v'$ and the path $u_v w'_v u'_v w_v$ to $G$, and we add the edge $w'_v u'_v$ to $M$.
Observe that $n' = \vert U \vert + \vert W \vert = O(n + k)$.

For each $a = v v' \in A$, note that $e_a= w_v u_{v'}$ is an element of $\overline{E}$ and put $c'(e_a) = c(a)$. 
Let $\overline{E}_A := \{e_a \mid a \in A\}$
be this set of edges. Every remaining edge $e \in \overline{E}$ 
has cost $c'(e) = 1+f(n+k) \cdot \sum_{a \in A} c(a)$ such that this edge is not contained 
in any $f(n')$-approximative solution.
This completes the construction of $G$, $c'$ and $M$. 
Observe that this transformation can be performed in polynomial time and that $M$ is indeed a perfect matching of $G$.
Additionally, there is a one-to-one correspondence between arcs in $A$ to edges 
in $\overline{E}_A$ as stated in Fact \ref{fact:arc-edge-correspondence}:
buying an arc in $A$ is equivalent to buying the corresponding edge in $\overline{E}_A$.

We now show that a feasible solution to \I can be transformed in polynomial time 
to a feasible solution of $\I'$ of the same cost. 
Let $X \subseteq A$ be a feasible solution to \I of cost $c(X)$ and let 
$X'$ be the corresponding edges to $X$ in $\overline{E}$. 
At first observe that $c(X) = c'(X')$. We now show that $X'$ is feasible to 
$\I'$.
By the one-to-one correspondence of arcs in $D$ and edges in $\overline{E}$, we have that
in $G+X'$ there is an alternating path $P_i$ from $u_{s_i}$ to $w_{t_i}$ for each $i \in [k]$. 
Thus, by adding the edge
$w_{t_i} u_{s_i}$ to $P_i$, we obtain an alternating cycle through $u_{t_i} w_{t_i}$ for each $i \in [k]$.
It follows that $X'$ is feasible.
Now let $X'$ be a solution to $\I'$ of cost $c'(X')$.
Let $X \subseteq A$ be the edges corresponding to the edge set $X' \cap E_A$.
Observe that $c(X) = c(X')$. We now show that $X$ is a feasible solution to \I.
As $X'$ is feasible to $\I'$, we have by Fact \ref{fact:kstrong} that every vertex
is contained in a directed cycle in $D(G+X, M)$. 
As a directed cycle through $u_{t_i}$ has to use the edge $u_{t_i} u_{s_i}$ (since
no terminal vertex appears more than once in \I),
the directed cycle in $D(G+X, M)$ has to go through $u_{s_i}$. 
This implies that there is a directed path from $s_i$ to $t_i$ in $D[X]$ for each $i \in [k]$
and therefore the feasibility of $X$.
Finally, as $n' = \vert U \vert + \vert W \vert = O(\vert V \vert + k)$, we have proved the first part of the proposition.

We now prove the second part: 
An $f(n)$- or $f(k)$-factor approximation algorithm for \dsnp implies an 
$(f(n)+1)$- or $(f(k) +1)$-factor approximation algorithm for \wonerma, respectively.
Let $\I$ be an instance of \wonerma with $G=(U + W, E)$ and $c \in \mathbb{Z}^{\overline{E}}_{\geq 0}$.
We set $c^* \in \mathbb{Z}^{E \cup \overline{E}}_{\geq 0}$
to $c^*(e) = c(e)$ if $e \in \overline{E}$ and $c(e) = 0$, otherwise.
Let $M = \{ u_1 w_1, \ldots, u_n w_n \}$ be any cost minimal perfect matching with respect to $c^*$, 
where $n = \vert U \vert = \vert W \vert$.
We construct the \dsnp instance $\I'$ with $D=(V', A)$, the terminal pairs $(s_1, t_1), \ldots, (s_k, t_k)$ 
and the cost function $c' \in \mathbb{Z}^A_{\geq 0}$ in the following way. 
We set $V' = V$ and add an arc $a_e = u w$ to $A$ if $e = u w \in M$ and add an arc 
$a_e = w u$ if $e = w u \in (E \cup \overline{E}) \setminus M$. 
In other words,  we direct the matching edges from $U$ to $W$ and the non-matching
edges from $W$ to $U$.
The terminal pairs are defined according to the matching, i.e.,
we let $s_i := w_i$ and $t_i := u_i$. Finally, for every $a \in A$, we let $c'(a) = 0$ if $e \in M$ and 
$c'(a_e) := c^*(e)$ otherwise. This completes the construction of $\I'$.

Let $X'$ be a feasible solution to $\I'$ of cost $c'(X')$.
Observe that by the chosen orientations of the arcs in $A$, 
any path from $w_i$ to $u_i$
in $X$ implies that there is an alternating path in the corresponding undirected graph with edge set $X$. 
Hence $X \cup M$ is feasible for $\I$.
Finally, as $M$ is a cost minimal matching with respect to $c^*$ and $k = O(n)$, 
we have that $M \cup X$
is an $(f(n) + 1)$- or $(f(k) + 1)$-factor approximation for \wonerma 
if $X'$ is an $f(n)$- or $f(k)$-factor approximation for \dsnp.
\end{proof}

\begin{proof}[Proof of Corollary~\ref{cor:wonerma:dsf:hardnessOfApproximation}]
  Observe that by the construction in the proof of Proposition \ref{prop:hardness:onerma-dsf}, we have that
  the number of vertices in the \wonerma instance is quadratic in the number of vertices from the \dsnp instance.
  Hence, by \cite{DBLP:conf/stoc/HalperinK03} and Proposition \ref{prop:hardness:onerma-dsf}, we have that for every $\varepsilon > 0$
  \wonerma does not admit a $\log^{2- \varepsilon}(n)$-factor approximation algorithm
  unless $\NP \subseteq \textsc{ZTIME}(n^{\polylog(n)})$.
\end{proof}

\begin{proof}[Proof of Lemma \ref{lemma:hardness:stars-and-paths}]
    The result follows in large parts from Lemma \ref{lemma:hardness:matching}. The main
    idea is that any instance of \wonerma on independent edges can be embedded in a
    sufficiently large member of $\mathcal{K}^*$ or $\mathcal{P}^*$.  More
    formally, consider an instance $\I = (G, M, c)$ of \wonerma, where $G$ 
    consists of independent edges. Let $(U, W)$ be any bipartition of $V(G)$.

    We first prove the statement for the class $\mathcal{K}^*$.  We construct an
    instance $\I' = (G', M', c')$ of \wonerma from \I, where $G' = K_{1,|M|+1}^*$. Let $G'$
    contain the independent edges $M$ and a path $P = v_1, v_2, v_3, v_4$, where
    $v_1, v_2, v_3, v_4$ are new vertices. For each $u \in U$, connect $v_2$ to $u$
    by an edge. Observe that $M' := M \cup \{v_1v_2, v_3v_4\}$ is a perfect
    matching of $G'$. The costs $c' \in \Z_{\geq 0}^{\overline{E(G')}}$ are given
    by
    \[
        c'_e :=
        \begin{cases}
            c_e, &   \text{if } e \in \overline{E(G)},\\
            0,   &   \text{if } e = v_1v_4, \\
            K,   &   \text{if } \text{otherwise},
        \end{cases}
    \]
    where $K$ is chosen such that no optimal solution contains an edge of weight $K$,
    for example, $K := |V(G')|\cdot \max_{e \in \overline{E}} c_e$.
    Since we may add $v_1v_4$ to any solution at no cost, we assume that it is
    present in any solution. Now, from the definition of $c'$ it follows that an
    optimal solution to $\I$ is also an optimal for $\I'$ and vice versa.

    It remains to prove the statement for the class $\{P_t^* \mid r \in \mathbb N\}$. In
    the following, let $n := |M|$.
    We construct an instance $\I''=(G'', M'', c'')$ of \wonerma from \I, where $G''
    = P_{2n}^*$. Let $G''$ contain the independent edges $M$ and join the
    vertices $U$ in any order by a path $P = v_1,u_1,v_2,u_2,\ldots,v_{n},u_{n}$,
    where $u_1, u_2, \ldots, u_{n} \in U$ and $v_1,v_2,\ldots,v_{n}$ are new
    vertices. Finally, for each $1 \leq i \leq n$, add a new vertex $v_i'$ to
    $G'$ and join it to $v_i$ by an edge. Let $M' := M \cup
    \{v_iv_i' \mid 1 \leq i \leq n\}$
    and let $c'' \in \Z_{\geq 0}^{\overline{E(G'')}}$ be given by
    \[
        c_e'' :=
        \begin{cases}
            c_e &   \text{if } e \in \overline{E(G)}\\
            0   &   \text{if } e = v_1'v_{n}\text{ or } e = v_iv_{i+1}',\, 1\leq i < n\\
            K   &   \text{otherwise},
        \end{cases}
    \]
    where $K$ is again chosen such that no optimal solution contains an edge of
    weight $K$, for example, $K := |V(G')|\cdot \max_{e \in \overline{E}} c_e$.
    By the choice of $c''$, we may assume that each edge in $M'' \setminus M$ is
    contained in an alternating cycle. Furthermore, since no optimal solution to
    $\I''$ connects $V(G'') \setminus V(G)$ to $V(G)$, we have that any optimal
    solution to $\I''$ is optimal for $\I$ and vice versa.
\end{proof}

\begin{proof}[Full Proof of Theorem~\ref{thm:weighted-main}]
	According to Lemma~\ref{lemma:hardness:stars-and-paths}, \wonerma is NP-hard if $\mathcal G$ completely contains the class $\mathcal{K} = \{K_{1,r}^* \mid r \in \mathbb N\}$ or the class $\mathcal{P} = \{P_t^* \mid r \in \mathbb N\}$. Assuming $\P\neq\NP$, this proves the \emph{only if} statement of the theorem.
	
	To see the \emph{if} statement, let us consider $r \in \mathbb N$ such that $\mathcal G$ does not contain $K_{1,r}^*$ or $P_r^*$.
	First we will reduce the problem to the case when $\mathcal G$ contains only trees.
	For this, let $\mathcal T$ be the class of all trees in $\mathcal G$ that admit a perfect matching.
	
	\setcounter{claim}{0}
        \begin{claim}
            There is a polynomial time reduction of \wonerma on $\mathcal G$ to \wonerma on $\mathcal T$.
            \label{clm:reductiom-to-trees2}
        \end{claim}
        \begin{proof}
            To see this, consider an input $(G,M,c)$ of \wonerma on $\mathcal G$, consisting of a bipartite graph $G \in \mathcal G$, a perfect matching $M$ of $G$, and costs $c$ of edges in the bipartite complement of $G$.
            We first compute a spanning tree $T$ of $G$ that contains all edges of $M$ using, e.g., Kruskal's algorithm.
            We extend the costs $c$ to all edges $e$ in the set $E(G) \setminus E(T)$ by setting $c_e =0$.

            Note that $(T,M,c)$ is an instance of \wonerma on $\mathcal T$.
            Moreover, for every optimal solution $S$ of the instance $(T,M,c)$, $S \setminus E(G)$ is an optimal solution of the instance $(G,M,c)$.
        \end{proof}
	
	We may hence restrict our attention to \wonerma on the class $\mathcal T$.
	As the next claim shows, the relevant trees contained in $\mathcal T$ have a bounded number of leaves.

        \begin{claim}
            There is some number $f(r)$ depending only on $r$ such that every tree in $\mathcal T$ has at most $f(r)$ many leaves.
            \label{clm:leaf-number2}
        \end{claim}
        \begin{proof}
            Let $T \in \mathcal T$ be arbitrary, and let $\ell$ be the number of leaves of $T$.
            Let us first show that the maximum degree of $T$ is bounded by $r$.
            Fix any perfect matching $M$ of $T$.
            Consider a vertex $v$ of $T$, and let $X$ be the set of all neighbors of $v$ together with their matching partners.
            Note that $T[X]$ is isomorphic to $K_{1,d(v)}^*$, where $d(v)$ denotes the degree of $v$.
            Since $\mathcal G$ is closed under taking connected minors, $K_{1,d(v)}^* \in \mathcal G$, and hence $d(v) < r$.

            Next, we show that the number of vertices of degree at least $3$ is bounded.
            Since the maximum degree of $T$ is bounded by $r$, the following holds for the number of leaves in $T$:
            \[
                \ell = 2 + \sum_{j=3}^r (j-2)|V_j|, \mbox{ where } V_j=\{v \in V(T) : d(v) = 3\}.
            \]
            The above formula is a standard graph theory exercise.
            As $r$ is constant, this implies $\sum_{j=3}^r |V_j| = \Omega(\ell)$.
            Again since $r$ is constant, there is a path in $T$ containing $\Omega( \log \ell)$ many vertices of degree at least $3$ in $T$.
            Let $T'$ be this path together with all vertices adjacent to it.

            Note that $P_t^*$ is a minor of $T'$ where $t+2$ is the number of vertices of degree at least $3$ on $T$.
            Since $\mathcal G$ is closed under connected minors and $P_r^* \notin \mathcal G$, we have $t < r$.
            Consequently, $t \in \Omega(\log \ell)$ implies that $\ell \le f(r)$ for some number $f(r)$ depending only on $r$.
        \end{proof}
	
	According to the above claims, there is a polynomial reduction of \wonerma on $\mathcal G$ to \wonerma on a class of trees with a bounded number of leaves.
	Hence, Lemma~\ref{lemma:weighted-trees} implies that \wonerma on $\mathcal G$ can be solved in polynomial time.
\end{proof}

\begin{proof}[Proof of Lemma \ref{lemma:weighted-trees}]
	Let $\I = (G, M, c)$ be an instance of \wonerma, where $G= (V, E) \in
	\mathcal{T}$ is a tree with at most $r$ leaves and a given bipartition $(U,W)$.
	Moreover, let $M$ be the unique perfect matching of $G$.  We say that
	an arc $xy$ is a \emph{shortcut} if there is an additional directed path
	from $x$ to $y$ in $D(G, M)$.
	
	\setcounter{claim}{0}	
	\begin{claim}\label{clm:no-shortcuts2}
		Let $L$ be an optimal solution to \I. Then we may assume that $D(G + L, M)$ contains no shortcut.
	\end{claim}
        \begin{proof}
            By Fact~\ref{fact:kstrong}, each strongly connected component of $D(G + L,
            M)$ is non-trivial. Suppose for a contradiction that $D(G + L, M)$ contains a
            shortcut arc $a$ and let $e \in \overline{E}$ be the edge corresponding to $a$.
            Then each strongly connected component of $D(G + (L - e), M)$ is non-trivial.
            Since the costs $c$ are non-negative, we conclude that $L-e$ is solution of
            weight at most $\OPT(\I)$.
        \end{proof}
	
	By Claim~\ref{clm:no-shortcuts2} we only need to augment edges that do not correspond to
	shortcuts in $D(G, M)$. So let $\tilde E \subseteq \overline{E}$ be the subset of edges that are useful
	for augmentation, that is,
	\[
	\tilde E := \{ uw \in \overline{E} \mid D(G + uw, M) \text{ has no shortcut} \} .
	\]
	For $F \subseteq E$, we denote by $F_{WU}$ the set of arcs obtained from $F$ by
	directing all edges from $W$ to $U$.  We construct a new directed graph $D'$
	on the vertices $V$ by directing all $M$-edges from $U$ to $W$ and making each
	edge in $E \setminus M$ bidirected.

        \begin{claim}
            Let $L' \subseteq \tilde E$. Then $G + L'$ is robust if and only if $D' + L'_{WU}$ is strongly connected.
            \label{clm:robust-if-strongly2}
        \end{claim}
        \begin{proof}
            First assume that $G + L'$ is robust and let $e = uw \in M$. Then $e$ is
            contained in some $M$-alternating cycle $C$ in $G + L'$. It is readily verified
            that there is a corresponding directed cycle in $D' + L'_{WU}$ containing the
            arc $uw$. Therefore, there is a path from $w$ to $u$ in $D'$.  Since the edges
            in $E \setminus M$ are undirected in $D'$, it follows that $D' + L'_{WU}$ is
            strongly connected. Now suppose that $D' + L_{WU}$ is strongly connected.
            Thus, each $M$-edge is contained in some cycle. Since $L' \subseteq \tilde E$,
            each $M$-edge is contained in an $M$-alternating cycle of $G + L'$,
            so $G + L'$ is robust.
        \end{proof}
	
	Using the two claims above we finish the proof of the lemma.
	By Claim~\ref{clm:robust-if-strongly2}, our task is to find a minimum-weight set $L' \subseteq \tilde E$,
	such that $D' + L'$ is strongly connected. For this purpose, we construct in
	polynomial time an instance $\I'$ of \dsnp with at most $r$ terminal pairs,
	such that from an optimal solution of $\I'$ we obtain an optimal solution of \I
	in a straight-forward manner. Since the number of terminals $r$ is constant, we can solve the \dsnp instance $\I'$ in polynomial using the algorithm from~\cite{feldman_ruhl_06} and obtain a solution of \I in polynomial time.
	
	The digraph of the instance $\I'$ is $D' + \tilde E_{WU}$ and the arc-costs
	$c'$ of $\I'$ are given as follows. For each arc $uw$ of $D' + \tilde E_{WU}$,
	let $c'_{uw}$ be
	\[
	c'_{uw} :=
	\begin{cases}
	c_{uw},    &    \text{if } uw \in \tilde E_{WU},\\
	0,        &    \text{otherwise} .
	\end{cases}
	\]
	The terminal pairs of $\I'$ are given as follows.  We run the algorithm
	\ET on $D(G, M)$ and obtain an arc-set $L$ such that $D(G, M) + L$ is strongly
	connected. By Fact~\ref{fact:minmax}, we have $|L| = \max \{ |\sources(D)|,
	|\sinks(D)| \} \leq r$. Each arc $a \in L$ corresponds to a pair of terminals
	we wish to connect. This completes the construction of $\I'$.
	
	We now show that
	optimal solutions of $\I'$ correspond to optimal solutions of $\I$. Let $L'$ be
	an optimal solution to $\I'$. That is, $D' + L'$ is strongly connected.  We
	may assume that $L'$ contains all arcs of $D'$, since each of them has weight
	zero. Since $L' \subseteq \tilde E_{WU}$, we invoke Claim~\ref{clm:robust-if-strongly2} and have that $G +
	\undirected(L')$ is strongly connected, where $\undirected(L')$ are the undirected edges
	corresponding to $L'$. Let $L \subseteq \tilde E$ such that $G+L$ is robust
	and assume that $c(L) < c(\undirected(L'))$. Then $L_{WU}$ is a solution of $\I'$ and
	$c'(L_{WU}) < c'(L')$. This contradicts the optimality of $L'$, so $\undirected(L')$
	is optimal for $\I$.
\end{proof}

\begin{lemma}\label{lemma:hardness:matching}
  \wonerpma on independent edges is \NP-hard.
\end{lemma}
\begin{proof}
    We reduce from \onerma, which was proved to be \NP-hard in
    Proposition~\ref{prop:hardness:onerma}. Let $\I = (G, M)$ be an instance of
    \onerma, where $G= (V, E)$.  We construct an instance $\I' = (G', M, c)$ of
    \wonerma as follows: Let $G' := (V, M)$ consist only of the edges from the
    perfect matching $M$.  Furthermore, let the costs $c \in \Z_{\geq
    0}^{\overline{E(G')}}$ be given by
    \[
        c_e :=
        \begin{cases}
            0,	&	\text{if } e \in E(G) \setminus M,\\
            1, &	\text{otherwise}.
        \end{cases}
    \] 
    Clearly, the construction can be performed in polynomial time. The solutions
    of $\I$ and $\I'$ are in 1-to-1 correspondence and the costs are preserved by
    the transformation.
\end{proof}
\section{Hardness Results}
\label{sec:hardness}

We show that the problem \onerma is \NP-hard, even on (bipartite) graphs of
maximum degree three. Furthermore, it is \NP-hard to find a $o(\log
n)$-approximate solution in polynomial time. The result mainly follows from the
results of~\cite{Bindewald:18} and an additional lemma.  Nevertheless, we give
the full proof here.

\begin{proposition}
	\label{prop:hardness:onerma}
	\onerma parameterized by the solution size is $\W[2]$-hard, even on graphs of maximum degree three.
\end{proposition}

\begin{proof}
	We give a parameterized reduction from \setcover, which is $\W[2]$-hard.
	Let $(X, \sets)$ be an instance of \setcover. We construct an instance $(G,
	M)$ of \onerma as follows. Let $d$ be the maximal cardinality of the sets in
	$\sets$. For each set $S \in \sets$, we add a cycle $C_S$ of length $2d$
	on the vertices $c_S^1, c_S^2, \ldots, c_S^{2d}$ and for each item $u \in
	X$, we add an edge $u_1u_2$ to $G$. For each $u \in X$ and $S \in \sets$,
	if $u \in S$, we join $u_1$ to $c_S^i$ by an edge, such that $i$ is odd and
	the vertex $c_S^i$ has maximum degree three. This is possible since $C_S$ has length
	$2d$. Finally, we add two vertices $t_1$ and $t_2$ to $G$, join them by an
	edge, and connect for each $u \in X$, $u_2$ to $t_1$. The matching $M$
	contains for each $S \in \sets$ the edges $c_S^1c_S^2,
	c_S^3c_S^4,\ldots,c_S^{2d-1}c_S^{2d}$ and for each $u \in X$ the edge
	$u_1u_2$, and also $t_1t_2$. It is readily verified that $M$ is a perfect
	matching of $G$. Let us choose the bipartition $(U, W)$ of $G$ such that
	$u_1 \in U$ for some $u \in X$. 
		
        \setcounter{claim}{0}
        \begin{claim}
            $C(D(G, M))$ contains a single sink $t_1$ and for each
            $S \in \sets$ its node-set $V(C_S)$ defines a strong source.
            \label{clm:single-sink-T-cycles-are-sources}
        \end{claim}
        \begin{proof}
            Clearly, the vertices of each cycle $C_S$ are in a strong component of
            $D(G, M)$. Observe that by the construction of $G$, any maximal
            $M$-alternating path that leaves a cycle $C_S$ terminates in $t_2$. It follows that
            $t_1$ is the only sink of $C(D(G, M))$.
            Moreover, no two distinct cycles $C_S$ and
            $C_{S'}$ are in the same strong component of $C(D(G, M))$.
        \end{proof}
	
	Let $L \subseteq \overline{E}$ be an optimal solution to $(G, M)$. By
	Fact~\ref{fact:sourcesink}, we can assume that $L$ connects sources to the unique
	sink of $C(D(G, M))$. Let 
	\[
            C_L := \{ S \in \sets \mid L \text{ connects $C_S$ to $t_1$} \}.
	\]
	
	Next we prove that $L$ is a solution of size $\ell$ if and only if $C_L$ is a solution of size $\ell$. 
	For the only if part, assume this is not true and let $u \in X$ be not covered
	by $C_L$. Thus none of the sets containing $u$ is contained in $C_L$, meaning that
	$L$ does not connect $t_1$ to a strong source that is a predecessor of $u_1$
	in $D(G, M)$ (as $L$ only connects $t_1$ to strong sources).
	Hence $u_1 u_2$ is not contained in an alternating cycle, a contradiction.
	For the if part, let $C_L$ be a cover of size $\ell$ and let $L$ be the 
	corresponding arcs in $D$. Assume $u_1$ is not contained in a strong component in $D(G+L, M)$. 
	As $L$ only connects strong sources to sinks, no predecessor of $u_1$ 
	has an edge to $t_1$. This is a contradiction to $C_L$ being a cover.
	
	We now describe how to reduce the degree of the constructed graph.
	Note that the only vertices with degree possibly greater than $3$ are 
	$t_1$ and $u_1$, $u \in X$. Both of them are in $U$.
	Consider a vertex $u \in U$ of degree at least $q > 3$ with its neighbors
	$w_1, \dots w_q$.
	We do not connect the vertices $w_i$, $1 \leq i \leq q$ directly to $u$.
	Instead we add a path $P = \{ u_1' w_1' u_2' w_2', \dots, u'_{q} = u \}$,
	where for $1 \leq i < q$ the edges $u_i' w_i'$ are matching edges.
	Instead of $w_i u$ we add the edges $w_i u_i'$ for $1 \leq i \leq q$.
	Observe that we still have the same properties as before
	but each vertex in $G$ has degree at most 3.
\end{proof}

\begin{proposition}
	\label{prop:hardness:inapprox}
	\onerma admits no polynomial time $o(\log n)$-factor approximation
	algorithm unless $\P = \NP$, where $n$ is the number of \unsafe edges
	of the input graph.
\end{proposition}

\begin{proof}
  Assume for a contradiction that there is a polynomial-time algorithm $A$ that
  computes a $f(n)$-approximate solution of \onerma, where $f(n) = o(\log n)$.
  Let $\I'=(X, \mathcal{S})$ be an instance of \setcover and construct from $\I'$ in
  polynomial time an instance $\I$ of \onerma as in the proof of Proposition
  \ref{prop:hardness:onerma}. We now also have that $\OPT(\I)= \OPT(\I')$ and $n=|X|$. 
  Applying algorithm $A$ on $\I$ yields a solution $L$ of cardinality
  at most $f(n)\cdot \OPT(\I)$.  Without loss of generality, we may assume that
  $L$ only connects sources and sinks due to Fact \ref{fact:sourcesink}. 
  We now set
  \[
    C_L := \{ S \in \sets \mid L \text{ connects $C_S$ to $t_1$} \}.
  \]
  By the same arguments as in the proof of Proposition
  \ref{prop:hardness:onerma}, we observe that $C_L$ is a feasible solution to
  $\I'$ of cardinality at most $f(n)\cdot \OPT(\I) = f(n')\cdot \OPT(\I')$.  This
  contradicts an inapproximability result of Dinur and Steurer for
  \setcover~\cite{DS:14}.
\end{proof}

\end{document}